# The discrete dipole approximation: an overview and recent developments

M.A. Yurkin[a,b*] and A.G. Hoekstra[a*]

[a] *Section Computational Science, Faculty of Science, University of Amsterdam, Kruislaan 403, 1098 SJ, Amsterdam, The Netherlands*

[b] *Institute of Chemical Kinetics and Combustion, Siberian Branch of the Russian Academy of Sciences, Institutskaya Str. 3, 630090, Novosibirsk, Russia*

## Abstract

We present a review of the Discrete Dipole Approximation (DDA), which is a general method to simulate light scattering by arbitrarily shaped particles. We put the method in historical context and discuss recent developments, taking the viewpoint of a general framework based on the integral equations for the electric field. We review both the theory of the DDA and its numerical aspects, the latter being of critical importance for any practical application of the method. Finally, the position of the DDA among other methods of light scattering simulation is shown and possible future developments are discussed.

---

* Corresponding authors. Tel.: +31-20-525-7462; fax: +31-20-525-7490.

*e-mail addresses*: myurkin@science.uva.nl, alfons@science.uva.nl

## 1 Introduction

The Discrete Dipole Approximation (DDA) is a general method to compute scattering and absorption of electromagnetic waves by particles of arbitrary geometry and composition. Initially the DDA was proposed by Purcell and Pennypacker (PP) [1], who replaced the scatterer by a set of point dipoles. These dipoles interact with each other and the incident field, giving rise to a system of linear equations, which is solved to obtain dipole polarizations. All the measured scattering quantities can be obtained from these polarizations. The DDA was further developed by Draine and coworkers [2-5], who popularized the method by developing a publicly available computer code DDSCAT [6]. Later it was shown that the DDA also can be derived from the integral equation for the electric field, which is discretized by dividing the scatterer into small cubical subvolumes. This derivation was apparently first performed by Goedecke and O'Brien [7] and further developed by others (see, for instance, [8-11]). It is important to note that the final equations, produced by both lines of derivation of the DDA are essentially the same. The only difference is that derivations based on the integral equations give more mathematical insight into the approximation, thus pointing at ways to improve the method, while the model based on point dipoles is physically clearer.

The DDA is called the coupled dipole method or approximation by some researchers [12,13]. There are also other methods, such as the volume integral equation formulation [14] and the Digitized Green's Function (DGF) [7], which were developed completely independently from PP. However, later they were shown to be equivalent to DDA [8,15]. In this review we will use the term DDA to refer to all such methods, since we describe them in terms of one general framework. However, it is difficult to separate unambiguously the DDA from other similar methods, based on the volume integral equations for the electromagnetic fields, such as a broad range of Method of Moments (MoM) with different bases and testing functions [16-19]. In our opinion, one fundamental aspect of the DDA is that the solution for the "physically meaningful" internal fields or their direct derivatives, e.g. polarization, plays an integral role in the process. In other words, any DDA formulation can be interpreted as replacing a scatterer by a set of interacting dipoles; this is further discussed in Section 2. An example of method that is not considered DDA is the MoM with higher-order hierarchical Legendre basis functions [17].

The DDA is a popular method in the light-scattering community and it has been reviewed by several authors. An extensive review by Draine and Flatau [4] covers almost all DDA developments up to 1994. A more recent review by Draine [5] mainly concerns applications and numerical considerations. DDA theory was discussed together with other methods for light scattering simulations in reviews by Wriedt [20], Chiappetta and Torresani [21], and Kahnert [15] and in books by Mishchenko *et al.* [22] and Tsang *et al.* [23]. Jones [24] placed the DDA in context of different methods with respect to particle characterization. However, many important DDA developments since 1994 are not mentioned in any of these manuscripts. Those that are mentioned are usually considered as side-steps, and are not placed into a general framework. Moreover, to the best of our knowledge numerical aspects of the DDA have never been reviewed extensively – each paper discusses only a few particular aspects. In this review we try to fill this gap.

A general framework is developed in Section 2 to ease the further discussion of different DDA models. This framework is based on the integral equation because it allows a uniform description of all the DDA development. However, connection to a physically clearer model of point dipoles is discussed throughout the section. The sources of errors in the DDA formulation are also discussed there.

In Section 3 the physical principles of the DDA are reviewed and results of different models are compared. In Subsection 3.1 different improvements of polarizabilities and interaction terms are reviewed from a theoretical point of view. Different expressions for $C_{abs}$ also are discussed. Comparison of simulation results using different formulations is given in Subsection 3.2. Subsection 3.3 covers the special case of a cluster of spheres that allows particular improvements and simplifications. In section 3.4 different significant modifications are reviewed, which do not fall completely into the general framework described in Section 2. Enhancements of the DDA for some special purposes also are discussed.

Different numerical aspects of the DDA are reviewed in Section 4. These are concerned primarily with solving very large systems of linear equations (Subsection 4.1). Subsection 4.2 describes the simplest iterative procedure to solve DDA linear system, which has a clear physical meaning. The special structure of the DDA interaction matrix for a rectangular grid and its application to decrease computational costs are described in Subsections 4.3 and 4.4 respectively. General methods to accelerate calculations, which do not require a rectangular grid, are discussed in Subsection 4.5. Subsection 4.6 covers special techniques to increase the efficiency of repeated calculations (e.g. in orientation averaging).

A numerical comparison of the DDA with other methods is reviewed in Section 5; its strong and weak points are discussed. Section 6 concludes the review and discusses future development of the DDA.

## 2 General framework

The $\exp(-\mathrm{i}\omega t)$ time dependence of all fields is assumed throughout this review. The scatterer is assumed dielectric but not magnetic (magnetic permittivity $\mu = 1$). The electric permittivity is assumed isotropic to simplify the derivations; however, extension to arbitrary dielectric tensors is straightforward.[1]

The general form of the integral equation governing the electric field inside the dielectric scatterer is the following [8,15]:

$$\mathbf{E}(\mathbf{r}) = \mathbf{E}^{inc}(\mathbf{r}) + \int\limits_{V \setminus V_0} d^3 r' \overline{\mathbf{G}}(\mathbf{r},\mathbf{r}')\chi(\mathbf{r}')\mathbf{E}(\mathbf{r}') + \mathbf{M}(V_0,\mathbf{r}) - \overline{\mathbf{L}}(\partial V_0,\mathbf{r})\chi(\mathbf{r})\mathbf{E}(\mathbf{r}), \tag{1}$$

where $\mathbf{E}^{inc}(\mathbf{r})$ and $\mathbf{E}(\mathbf{r})$ are the incident and total electric field at location $\mathbf{r}$; $\chi(\mathbf{r}) = (\varepsilon(\mathbf{r}) - 1)/4\pi$ is the susceptibility of the medium at point $\mathbf{r}$ ($\varepsilon(\mathbf{r})$ – relative permittivity). $V$ is the volume of the particle, i.e., the volume that contains all points where the susceptibility is not zero. $V_0$ is a smaller volume such that $V_0 \subset V$, $\mathbf{r} \in V_0 \setminus \partial V_0$. $\overline{\mathbf{G}}(\mathbf{r},\mathbf{r}')$ is the free space dyadic Green's function, defined as

$$\overline{\mathbf{G}}(\mathbf{r},\mathbf{r}') = \left(k^2\overline{\mathbf{I}} + \hat{\nabla}\hat{\nabla}\right)\frac{\exp(\mathrm{i}kR)}{R} = \frac{\exp(\mathrm{i}kR)}{R}\left[k^2\left(\overline{\mathbf{I}} - \frac{\hat{R}\hat{R}}{R^2}\right) - \frac{1-\mathrm{i}kR}{R^2}\left(\overline{\mathbf{I}} - 3\frac{\hat{R}\hat{R}}{R^2}\right)\right], \tag{2}$$

where $\overline{\mathbf{I}}$ is the identity dyadic, $k = \omega/c$ is the free space wave vector, $\mathbf{R} = \mathbf{r} - \mathbf{r}'$, $R = |\mathbf{R}|$, and $\hat{R}\hat{R}$ is a dyadic defined as $\hat{R}\hat{R}_{\mu\nu} = R_\mu R_\nu$ ($\mu$ and $\nu$ are Cartesian components of the vector or tensor). $\mathbf{M}$ is the following integral associated with the finiteness of the exclusion volume $V_0$

$$\mathbf{M}(V_0,\mathbf{r}) = \int_{V_0} d^3r' \left(\overline{\mathbf{G}}(\mathbf{r},\mathbf{r}')\chi(\mathbf{r}')\mathbf{E}(\mathbf{r}') - \overline{\mathbf{G}}^s(\mathbf{r},\mathbf{r}')\chi(\mathbf{r})\mathbf{E}(\mathbf{r})\right), \tag{3}$$

where $\overline{\mathbf{G}}^s(\mathbf{r},\mathbf{r}')$ is the static limit ($k \to 0$) of $\overline{\mathbf{G}}(\mathbf{r},\mathbf{r}')$:

$$\overline{\mathbf{G}}^s(\mathbf{r},\mathbf{r}') = \hat{\nabla}\hat{\nabla}\frac{1}{R} = -\frac{1}{R^3}\left(\overline{\mathbf{I}} - 3\frac{\hat{R}\hat{R}}{R^2}\right). \tag{4}$$

[1] In most formulae scalar values can be replaced directly by tensors, but there are exceptions. Extensions of DDA to optically anisotropic scatterers are discussed in Section 3.4.

$\overline{\mathbf{L}}$ is the so-called self-term dyadic:

$$\overline{\mathbf{L}}(\partial V_0,\mathbf{r}) = -\oint_{\partial V_0} d^2r' \frac{\hat{n}'\hat{R}}{R^3}, \tag{5}$$

where $\hat{n}'$ is an external normal to the surface $\partial V_0$ at point $\mathbf{r}'$. $\overline{\mathbf{L}}$ is always a real, symmetric dyadic with trace equal to $4\pi$ [25]. It is important to note that $\overline{\mathbf{L}}$ does not depend on the size of the volume $V_0$, but only on its shape (and location of the point $\mathbf{r}$ inside it). On the contrary, $\mathbf{M}$ does depend on the size of the volume, moreover it approaches zero when the size of the volume decreases [8] (if both $\chi(\mathbf{r})$ and $\mathbf{E}(\mathbf{r})$ are continuous inside $V_0$).

When deriving Eq. (1) the singularity of the Green's function has been treated explicitly, therefore it is preferable to the commonly used formulation [8,15]:

$$\mathbf{E}(\mathbf{r}) = \mathbf{E}^{inc}(\mathbf{r}) + \int_V d^3r' \overline{\mathbf{G}}(\mathbf{r},\mathbf{r}')\chi(\mathbf{r}')\mathbf{E}(\mathbf{r}'). \tag{6}$$

Moreover, Yanghjian noted [25] that there exist several methods for treating the singularity in Eq. (6) leading to different results. He also proved that the derivation of Eq. (6) is false in the vicinity of the singularity of $\overline{\mathbf{G}}(\mathbf{r},\mathbf{r}')$. Hence it can be considered correct only if the singularity is then treated in a way similar to that of Lakhtakia [8], resulting in the correct Eq. (1).

Discretization of Eq. (1) is done in the following way [15]. Let $V = \bigcup_{i=1}^{N} V_i$ , $V_i \cap V_j = \emptyset$ for $i \neq j$; $N$ denotes number of subvolumes.[2] Although the formulation is applicable to any set of subvolumes $V_i$, in most applications standard (equal) cells are used. Then the shape of the scatterer cannot always be described exactly by such standard cells. Hence, the discretization may be only approximately correct. Assuming $\mathbf{r} \in V_i$ and choosing $V_0 = V_i$, Eq. (1) can be rewritten as

$$\mathbf{E}(\mathbf{r}) = \mathbf{E}^{inc}(\mathbf{r}) + \sum_{j \neq i} \int_{V_j} d^3 r' \overline{\mathbf{G}}(\mathbf{r},\mathbf{r}')\chi(\mathbf{r}')\mathbf{E}(\mathbf{r}') + \mathbf{M}(V_i,\mathbf{r}) - \overline{\mathbf{L}}(\partial V_i,\mathbf{r})\chi(\mathbf{r})\mathbf{E}(\mathbf{r}) . \tag{7}$$

The set of Eq. (7) (for all $i$) is exact. Further, one fixed point $\mathbf{r}_i$ inside each $V_i$ (its center) is chosen and $\mathbf{r} = \mathbf{r}_i$ is set. In many cases the following assumptions can be made:

$$\int_{V_j} d^3 r' \overline{\mathbf{G}}(\mathbf{r}_i,\mathbf{r}')\chi(\mathbf{r}')\mathbf{E}(\mathbf{r}') = V_j \overline{\mathbf{G}}_{ij}\chi(\mathbf{r}_j)\mathbf{E}(\mathbf{r}_j) , \tag{8}$$

$$\mathbf{M}(V_i,\mathbf{r}_i) = \overline{\mathbf{M}}_i \chi(\mathbf{r}_i)\mathbf{E}(\mathbf{r}_i) , \tag{9}$$

which state that integrals in Eq. (7) linearly depend upon the values of $\chi$ and $\mathbf{E}$ at point $\mathbf{r}_i$. Eq. (7) can then be rewritten as

$$\mathbf{E}_i = \mathbf{E}_i^{inc} + \sum_{j \neq i} \overline{\mathbf{G}}_{ij} V_j \chi_j \mathbf{E}_j + \left(\overline{\mathbf{M}}_i - \overline{\mathbf{L}}_i\right)\chi_i \mathbf{E}_i , \tag{10}$$

where $\mathbf{E}_i = \mathbf{E}(\mathbf{r}_i)$, $\mathbf{E}_i^{inc} = \mathbf{E}^{inc}(\mathbf{r}_i)$, $\chi_i = \chi(\mathbf{r}_i)$, $\overline{\mathbf{L}}_i = \overline{\mathbf{L}}(\partial V_i,\mathbf{r}_i)$.

The usual approximation [15] is to consider $\mathbf{E}$ and $\chi$ constant inside each subvolume:

$$\mathbf{E}(\mathbf{r}) = \mathbf{E}_i, \ \chi(\mathbf{r}) = \chi_i \ \text{ for } \ \mathbf{r} \in V_i , \tag{11}$$

which automatically implies Eqs. (8), (9) and

$$\overline{\mathbf{M}}_i^{(0)} = \int_{V_i} d^3 r' \left(\overline{\mathbf{G}}(\mathbf{r}_i,\mathbf{r}') - \overline{\mathbf{G}}^s(\mathbf{r}_i,\mathbf{r}')\right), \tag{12}$$

$$\overline{\mathbf{G}}_{ij}^{(0)} = \frac{1}{V_j} \int_{V_j} d^3 r' \overline{\mathbf{G}}(\mathbf{r}_i,\mathbf{r}') . \tag{13}$$

Superscript (0) denotes approximate values of the dyadics. A further approximation, which is used in almost all formulations of the DDA, including e.g. [8], is

$$\overline{\mathbf{G}}_{ij}^{(0)} = \overline{\mathbf{G}}(\mathbf{r}_i,\mathbf{r}_j) . \tag{14}$$

This assumption is made implicitly by all formulations that start by replacing the scatterer with a set of point dipoles. It is important to note that Eq. (10) and derivations resulting from it require weaker assumptions (Eqs. (8), (9)) than imposed by Eq. (11) and, moreover,

---

[2] In the framework of the DDA we usually call a subvolume a dipole.

Eq. (14). It is possible to formulate the DDA based on Eq. (10), e.g. the Peltoniemi formulation [26] that is described in Section 3.1. We postulate Eq. (10) as a distinctive feature of the DDA, i.e. a method is called the DDA if and only if its main equation is equivalent to Eq. (10) with any $V_i$, $\chi_i$, $\overline{\mathbf{M}}_i$, $\overline{\mathbf{L}}_i$, and $\overline{\mathbf{G}}_{ij}$.

Kahnert [15] distinguished the DDA from the MoM (Method of Moments) by the fact that the MoM solves directly Eq. (10) for unknown $\mathbf{E}_i$, while the DDA seeks not the total, but the exciting electric fields

$$\mathbf{E}_i^{exc} = \left(\overline{\mathbf{I}} + \left(\overline{\mathbf{L}}_i - \overline{\mathbf{M}}_i\right)\chi_i\right)\mathbf{E}_i = \mathbf{E}_i - \mathbf{E}_i^{self}, \tag{15}$$

$$\mathbf{E}_i^{self} = \left(\overline{\mathbf{M}}_i - \overline{\mathbf{L}}_i\right)\chi_i\mathbf{E}_i, \tag{16}$$

where $\mathbf{E}_i^{self}$ is the field induced by the subvolume on itself. Eq. (10) is then equivalent to

$$\mathbf{E}_i^{inc} = \mathbf{E}_i^{exc} - \sum_{j\neq i}\overline{\mathbf{G}}_{ij}\overline{\boldsymbol{\alpha}}_j\mathbf{E}_j^{exc}, \tag{17}$$

where $\overline{\boldsymbol{\alpha}}_i$ is the polarizability tensor defined as

$$\overline{\boldsymbol{\alpha}}_i = V_i\chi_i\left(\overline{\mathbf{I}} + \left(\overline{\mathbf{L}}_i - \overline{\mathbf{M}}_i\right)\chi_i\right)^{-1}. \tag{18}$$

However, an alternative formulation of the DDA exists [4] seeking a solution for unknown polarizations $\mathbf{P}_i$:

$$\mathbf{P}_i = \overline{\boldsymbol{\alpha}}_i\mathbf{E}_i^{exc} = V_i\chi_i\mathbf{E}_i, \tag{19}$$

$$\mathbf{E}_i^{inc} = \overline{\boldsymbol{\alpha}}_i^{-1}\mathbf{P}_i - \sum_{j\neq i}\overline{\mathbf{G}}_{ij}\mathbf{P}_j. \tag{20}$$

It is important to note that $\mathbf{P}_i$, defined by Eq. (19), is only an approximation to the polarization of the subvolume $V_i$. This approximation is exact only under the assumption of Eq. (11), while the formulation itself does not require it. The formulation, using Eq. (20), can be thought as an intermediary between the DDA and the MoM as classified by Kahnert [15], therefore revealing complete equivalence of these two formulations. The special structure of the matrix $\overline{\mathbf{G}}_{ij}$ makes Eq. (20) preferable over Eqs. (10), (17) to find a numerical solution. This is discussed in Section 4.

Lakhtakia [8] classified strong and weak forms of the DDA as those accounting for or neglecting $\overline{\mathbf{M}}_i$ respectively. The weak form approaches the strong form when the size of the cell decreases, because $\overline{\mathbf{M}}_i$ approaches zero. For a cubical cell $V_i$ and with $\mathbf{r}_i$ located at the center of the cell, $\overline{\mathbf{L}}_i$ can be calculated analytically yielding [25]

$$\overline{\mathbf{L}}_i = \frac{4\pi}{3}\overline{\mathbf{I}}. \tag{21}$$

Using Eq. (18), this results in the well-known Clausius-Mossotti (CM) polarizability (used originally by Purcell and Pennypacker [1]) for the weak form of the DDA:

$$\bar{\boldsymbol{\alpha}}_i = \bar{\mathbf{I}}\alpha_i^{CM} = \bar{\mathbf{I}}d^3 \frac{3}{4\pi}\frac{\varepsilon_i - 1}{\varepsilon_i + 2}, \tag{22}$$

where $\varepsilon_i = \varepsilon(\mathbf{r}_i)$, and $d$ is the size of the cubical cell.

After the internal electric fields are determined, the scattered fields and cross sections can be calculated. The scattered fields are obtained by taking the limit $r \to \infty$ of the integral in Eq. (1) (see e.g. [7]):

$$\mathbf{E}^{sca}(\mathbf{r}) = \frac{\exp(\mathrm{i}kr)}{-\mathrm{i}kr}\mathbf{F}(\mathbf{n}), \tag{23}$$

where $\mathbf{n} = \mathbf{r}/r$ is the unit vector in the scattering direction, and $\mathbf{F}$ is the scattering amplitude:

$$\mathbf{F}(\mathbf{n}) = -\mathrm{i}k^3(\bar{\mathbf{I}} - \hat{n}\hat{n})\sum_i \int_{V_i} d^3r' \exp(-\mathrm{i}k\mathbf{r}'\cdot\mathbf{n})\chi(\mathbf{r}')\mathbf{E}(\mathbf{r}'). \tag{24}$$

All other differential scattering properties, such as amplitude and Mueller scattering matrices, and asymmetry parameter $\langle\cos\theta\rangle$ can be derived from $\mathbf{F}(\mathbf{n})$, calculated for two incident polarizations [27]. Radiation forces also can be calculated [28-30]. Consider an incident polarized plane wave[3]

$$\mathbf{E}^{inc}(\mathbf{r}) = \mathbf{e}^0 \exp(\mathrm{i}\mathbf{k}\cdot\mathbf{r}), \tag{25}$$

where $\mathbf{k} = k\mathbf{a}$, $\mathbf{a}$ is the incident direction, and $|\mathbf{e}^0| = 1$. The scattering cross section $C_{sca}$ is [27]

$$C_{sca} = \frac{1}{k^2}\oint d\Omega |\mathbf{F}(\mathbf{n})|^2. \tag{26}$$

Absorption and extinction cross sections ($C_{abs}$, $C_{ext}$) are derived [7,14] directly from the internal fields:

$$C_{abs} = 4\pi k\sum_i \int_{V_i} d^3r' \operatorname{Im}\left(\chi(\mathbf{r}')\right)|\mathbf{E}(\mathbf{r}')|^2, \tag{27}$$

$$C_{ext} = 4\pi k\sum_i \int_{V_i} d^3r' \operatorname{Im}\left(\chi(\mathbf{r}')\mathbf{E}(\mathbf{r}')\cdot\left[\mathbf{E}^{inc}(\mathbf{r}')\right]^*\right) = \frac{4\pi}{k^2}\operatorname{Re}\left(\mathbf{F}(\mathbf{a})\cdot\mathbf{e}^{0*}\right), \tag{28}$$

where * denotes a complex conjugate. Conservation of energy necessitates that

$$C_{sca} = C_{ext} - C_{abs}. \tag{29}$$

However, as was noted by Draine [2], use of Eq. (29) for evaluation of $C_{sca}$ can lead to larger errors than Eq. (26), especially when $C_{abs} \gg C_{sca}$.

---

[3] DDA can be used for any incident wave, e.g. Gaussian beams [31]; however, we do not discuss this here.

The easiest way to express Eqs. (24) and (27) in terms of the internal fields in the subvolumes centers is to assume Eq. (11), yielding

$$\mathbf{F}^{(0)}(\mathbf{n}) = -\mathrm{i}k^3(\bar{\mathbf{I}} - \hat{n}\hat{n})\sum_i \chi_i \mathbf{E}_i \int_{V_i} d^3r' \exp(-\mathrm{i}k\mathbf{r}'\cdot\mathbf{n}), \tag{30}$$

$$C_{abs}^{(0)} = 4\pi k \sum_i V_i \operatorname{Im}(\chi_i)\left|\mathbf{E}_i\right|^2 = 4\pi k \sum_i \operatorname{Im}(\mathbf{P}_i \mathbf{E}_i^*). \tag{31}$$

Further approximation of Eq. (30), leaving only the lowest order expansion of the exponent around $\mathbf{r}_i$, leads to

$$\mathbf{F}^{(0)}(\mathbf{n}) = -\mathrm{i}k^3(\bar{\mathbf{I}} - \hat{n}\hat{n})\sum_i \mathbf{P}_i \exp(-\mathrm{i}k\mathbf{r}_i\cdot\mathbf{n}), \tag{32}$$

which together with Eq. (28), leads to

$$C_{ext}^{(0)} = 4\pi k \sum_i \operatorname{Im}\left(\mathbf{P}_i \cdot \mathbf{E}_i^{inc*}\right). \tag{33}$$

Eqs. (32) and (33) are identical to those used by Purcell and Pennypacker [1] and then by Draine [2], while expressions for $C_{abs}$ (compared to Eq. (31)) are slightly different. These differences are discussed in Subsection 3.1. Unfortunately, many researchers do not specify explicitly how the scattering quantities are obtained from the computed internal fields or polarizations. Those who do usually use Draine's prescription (Eqs. (26), (32), (33), and (35)).

Errors of the formulation can be classified as associated with the finite cell size $d$ (discretization errors), and with approximating the particle shape with a set of standard cells, e.g. cubical (shape errors). Discretization errors result from considering $\mathbf{E}$ constant inside each cell and the approximate evaluation of $\overline{\mathbf{M}}_i$ and $\overline{\mathbf{G}}_{ij}$. Shape errors also can be considered as resulting from the assumption of constant $\chi$ and $\mathbf{E}$ inside bordering cells, which is false since the edge of the particle crosses these cells. On the other hand, shape errors can be viewed as a difference of the results for the exact particle shape and for that comprised of the set of standard cells. Both errors approach zero when $N \to \infty$, while the geometry of the scatterer and parameters of the incident field are fixed. However, the same does not apply if $kd \to 0$ while $N$ is fixed, i.e. the DDA is not exact in the long-wavelength limit. Moreover, both errors are sensitive to the size of the scatterer in the resonance region (see discussion in Subsection 3.2). The behavior of these errors was studied by Yurkin *et al.* [32].

# 3 Various DDA models

## 3.1 Theoretical base of the DDA

Since the original manuscript by Purcell and Pennypacker [1], many attempts have been made to improve the DDA. The first stage (1988-1993) of these improvements was reviewed by Draine and Flatau [4]. It has been noted [2] that Eq. (22) does not satisfy energy conservation, and results obtained using this formulation do not satisfy the optical theorem. Based on the well-known [33] "radiative reaction" (RR) electric field, a correction to the polarizability for a *finite* dipole was added [2]:

$$\alpha^{RR} = \frac{\alpha^{CM}}{1-(2/3)\mathrm{i}k^3\alpha^{CM}}. \tag{34}$$

Draine [2] also proposed the following expression for the absorption cross section:

$$C_{abs}^{(0)} = 4\pi k\sum_i \left[\mathrm{Im}(\mathbf{P}_i \cdot \mathbf{E}_i^{exc*}) - (2/3)k^3\mathbf{P}_i \cdot \mathbf{P}_i^*\right], \tag{35}$$

derived from Eq. (29) applied to a single *point* dipole. The PP formulation uses Eq. (35) without the second part. It can be verified that Eq. (35) results in zero absorption for any scatterer if the polarizability is of the following form:

$$\overline{\boldsymbol{\alpha}}_i^{-1} = \overline{\mathbf{A}}_i - (2/3)\mathrm{i}k^3\overline{\mathbf{I}},\ \overline{\mathbf{A}}_i = \overline{\mathbf{A}}_i^H, \tag{36}$$

where $H$ denotes the conjugate transpose of a tensor. For real refractive index $m$, RR and all other expressions specified below result in $\overline{\boldsymbol{\alpha}}$ satisfying Eq. (36), which makes Eq. (35) clearly favorable over e.g. the PP formulation. It must be noted however that the original PP formulation, where CM polarizability was used, also results in zero absorption for real $m$.

The correction in Eq. (34) is $\mathrm{O}\left((kd)^3\right)$. Several other corrections of $\mathrm{O}\left((kd)^2\right)$ have been proposed. The first one was proposed by Goedecke and O'Brien [7] and independently in two other manuscripts [34,35]. They started from Eqs. (10)-(12) and used the following simplifying fact for a cubical cell (also valid for spherical cells), resulting from symmetry:

$$\int_{cube} d^3R f(R)\frac{\hat{R}\hat{R}}{R^2} = \int_{cube} d^3R f(R)\frac{1}{3}\overline{\mathbf{I}}, \tag{37}$$

where the origin is in the center of the cube. Eq. (37) is valid for any $f(R)$ that has a singularity of less than third order for $R \to 0$, i.e. the integrals on both sides are defined. They obtained

$$\overline{\mathbf{M}}_i^{(0)} = \overline{\mathbf{I}}\frac{2}{3}k^2\int_{cube} d^3R\frac{\exp(\mathrm{i}kR)}{R}. \tag{38}$$

By expanding $\exp(\mathrm{i}kR)$ in Taylor series one can obtain

$$\overline{\mathbf{M}}_i^{(0)} = \overline{\mathbf{I}}\frac{2}{3}k^2\left(\int_{cube}\frac{d^3R}{R} + \mathrm{i}kd^3 + \mathrm{O}\left(k^2d^4\right)\right). \tag{39}$$

The remaining integral was evaluated by approximating the cube by a volume-equivalent sphere, resulting in

$$\overline{\mathbf{M}}_i^{(0)} = \overline{\mathbf{I}}\left(b_1^{DGF}(kd)^2 + (2/3)\mathrm{i}(kd)^3 + \mathrm{O}\left((kd)^4\right)\right), \tag{40}$$

$$b_1^{DGF} = (4\pi/3)^{1/3} \approx 1.611992. \tag{41}$$

An exact evaluation, obtained without expanding the exponent, of Eq. (38) for the equivolume sphere with radius $a = d(3/4\pi)^{1/3}$ was performed by Livensay and Chen [36] and implemented into the DGF formulation of the DDA by Hage and Greenberg [14,35] and later Lakhtakia [37]:

$$\overline{\mathbf{M}}_i^{(0)} = (8\pi/3)\overline{\mathbf{I}}\left[(1-\mathrm{i}ka)\exp(\mathrm{i}ka)-1\right]. \tag{42}$$

In terms of the first two orders of expansion, this yields an identical result as Eq. (40). Finally the polarizability is obtained as

$$\alpha^{DGF} = \frac{\alpha^{CM}}{1-\left(\alpha^{CM}/d^3\right)\left(b_1^{DGF}(kd)^2+(2/3)\mathrm{i}(kd)^3\right)}. \tag{43}$$

We denote the method based on Eq. (42) as LAK. Differences between LAK and DGF should be noticeable only for large values of *kd*.

Dungey and Bohren [38], using results by Doyle [39], proposed the following treatment of the polarizability. First, each cubic cell is replaced by the inscribed sphere that is called a dipolar subunit with a higher relative electric permittivity $\varepsilon_s$ as determined by the Maxwell-Garnett effective medium theory [27]:

$$f\frac{\varepsilon_s-1}{\varepsilon_s+2} = \frac{\varepsilon-1}{\varepsilon+2}, \tag{44}$$

where $f = \pi/6$ is the volume filling factor. Other effective medium theories also may be used [40]. Next, the dipole moment of the equivalent sphere is determined using the Mie theory, and the polarizability is defined as [39]

$$\alpha^M = i\frac{3}{2k^3}\alpha_1, \tag{45}$$

where $\alpha_1$ is the electric dipole coefficient from the Mie theory (see e.g. [41]):

$$\alpha_1 = \frac{m_s\psi_1(m_s x_s)\psi_1'(x_s)-\psi_1(x_s)\psi_1'(m_s x_s)}{m_s\psi_1(m_s x_s)\xi_1'(x_s)-\xi_1(x_s)\psi_1'(m_s x_s)}, \tag{46}$$

where $\psi$, $\xi$ are Riccati-Bessel functions; $x_s = kd/2$ and $m_s = \sqrt{\varepsilon_s}$ are the size parameter and the relative refractive index of the equivalent sphere. We denote this formulation for the polarizability as the $a_1$-term method (note that this terminology was introduced later [42]). It has the particular property that $\alpha^M/\alpha^{CM} \to \mathrm{const} \neq 1$ when $m \to 1$, contrary to all other polarization prescription, for which this ratio approaches 1. It should be noted that the Mie theory is based on the assumption that the external electric field is a plane wave. In most applications of the DDA this is true for the incident electric field, but not for the field created by other subvolumes. Therefore the $a_1$-term method is expected to be correct only for very small cell size. Hence it is not clear whether this method has advantages even compared to CM. On the other hand, this method may be more justified for clusters of small spheres, where each sphere can be considered as a dipole (see Subsection 3.3).

Draine and Goodman [3] pointed out that considering electric fields constant for evaluating integrals over a cell introduces errors of order $\mathrm{O}\left((kd)^2\right)$. This represents a problem for many polarizability corrections, based on integral equations. Draine and Goodman approached this problem from a different angle. They determined the optimal polarizability in the sense that an infinite lattice of point dipoles with such polarizability would lead to the same propagation of plane waves[4] as in a medium with a given refractive index. This polarizability was called LDR (Lattice Dispersion Relation) and is, as expected, CM plus high-order corrections. These corrections in turn depend on the direction of propagation **a** and the polarization of the incident field $\mathbf{e}^0$:

$$\alpha^{LDR} = \frac{\alpha^{CM}}{1-\left(\alpha^{CM}/d^3\right)\left[\left(b_1^{LDR}+b_2^{LDR}m^2+b_3^{LDR}m^2 S\right)(kd)^2+(2/3)\mathrm{i}(kd)^3\right]}, \tag{47}$$

$$b_1^{LDR} \approx 1.8915316,\ b_2^{LDR} \approx -0.1648469,\ b_3^{LDR} \approx 1.7700004\ , \tag{48}$$

$$S = \sum_{\mu}\left(a_\mu e_\mu^0\right)^2 . \tag{49}$$

We use a reverse sign convention in the denominator of Eq. (47) and the LDR coefficients as compared to the original paper [3].

Recently it has been shown [43] that the LDR derivation is not completely accurate, since the resulting dipole moment does not satisfy the transversality condition, for which a correction was proposed. This Corrected LDR (CLDR) differs principally in the fact that the polarizability tensor can not be made isotropic but only diagonal [43], though not dependent on the incident polarization:

$$\alpha_{\mu\nu}^{CLDR} = \frac{\alpha^{CM}\delta_{\mu\nu}}{1-\left(\alpha^{CM}/d^3\right)\left[\left(b_1^{LDR}+b_2^{LDR}m^2+b_3^{LDR}m^2 a_\mu^2\right)(kd)^2+(2/3)\mathrm{i}(kd)^3\right]}. \tag{50}$$

Another flaw of LDR is that it is evidently not correct for dipoles near the particle surface. However, it is not clear how to evaluate the effect of these mistreated surface dipoles on the overall results, e.g. on the scattering cross section.

Further improvement of the DDA was initiated by Peltoniemi [26] (PEL) who showed that the term $\mathbf{M}(V_i)$ in Eq. (7) can be evaluated exactly up to the third order of $kd$ by expanding the term $\chi(\mathbf{r}')\mathbf{E}(\mathbf{r}')$ under the integral in a Taylor series over the point $\mathbf{r}'=\mathbf{r}_i$, yielding

[4] with certain direction of propagation and polarization state.

$$
\begin{aligned}
M_{\mu}(V_i) = \sum_{\nu} M^{(0)}_{i,\mu\nu}\chi E_{\nu} + \frac{1}{2}\int_{V_i} d^3R\frac{\exp(\mathrm{i}kR)}{R^3}\left(k^2R^2 + \mathrm{i}kR - 1\right)\sum_{\nu} R_{\nu}^2\partial_{\nu}^2\chi E_{\mu} \\
-\frac{1}{2}\int_{V_i} d^3R\frac{\exp(\mathrm{i}kR)}{R^5}\left(k^2R^2 + 3\mathrm{i}kR - 3\right)\sum_{\nu\rho\tau} R_{\mu}R_{\nu}R_{\rho}R_{\tau}\partial_{\rho}\partial_{\tau}\chi E_{\nu} + \mathrm{O}\left((kd)^4\chi E\right),
\end{aligned}
\tag{51}
$$

where $\chi$, **E** and their derivatives are all considered at the point $\mathbf{r}_i$. Eq. (51) is correct up to the third order of $kd$ since the third term in the Taylor series vanishes because of symmetry. For spherical $V_i$ of radius $a$, the integrals can be evaluated exactly [26] in a way similar to obtaining Eq. (42), but only terms of less than fourth order of $kd$ are significant, which results in

$$
\mathbf{M}(V_i) = \frac{4\pi}{3}\left(\left((ka)^2 + \frac{2}{3}\mathrm{i}(ka)^3\right)\chi\mathbf{E} - a^2\left(\frac{1}{10}\nabla^2\chi\mathbf{E} - \frac{3}{10}\nabla(\nabla\cdot\chi\mathbf{E})\right)\right) + \mathrm{O}\left((ka)^4\chi E\right). \tag{52}
$$

If $\chi$ is constant inside the cell then the Maxwell equations state that

$$
\nabla^2\mathbf{E} = -m^2k^2\mathbf{E},\ \nabla\cdot\mathbf{E} = 0. \tag{53}
$$

Hence Eq. (9) is valid up to the third order of $ka$ and

$$
\overline{\mathbf{M}}_i = (4\pi/3)\bar{\mathbf{I}}\left[\left(1 + (1/10)m^2\right)(ka)^2 + (2/3)\mathrm{i}(ka)^3\right]. \tag{54}
$$

Piller and Martin [44] proposed using sampling theory to evaluate the integrals in Eq. (1). The electric field and the susceptibility is sampled:

$$
\chi(\mathbf{r}')\mathbf{E}(\mathbf{r}') = d^3\sum_i h^r(\mathbf{r}' - \mathbf{r}_i)\chi(\mathbf{r}_i)\mathbf{E}(\mathbf{r}_i), \tag{55}
$$

where $h^r(\mathbf{r})$ is the impulse response function of an antialiasing filter defined as

$$
h^r(\mathbf{r}) = \frac{\sin(qr) - qr\cos(qr)}{2\pi^2r^3}, \tag{56}
$$

where $q = \pi/d$. Eq. (1) is then transformed to Eq. (10) with the so-called filtered Green's function, defined as

$$
\overline{\mathbf{G}}_{ij} = \int_{\mathbf{R}^3/V_0} d^3r'\overline{\mathbf{G}}(\mathbf{r}_i,\mathbf{r}')h^r(\mathbf{r}' - \mathbf{r}_j). \tag{57}
$$

Eq. (57) can be viewed as a generalization of Eq. (13). The latter is obtained if a pulse function is considered instead of $h^r$. The integral in Eq. (57) is evaluated analytically [44], taking $V_0$ to be infinitesimally small. The filtered Green's function does not have a singularity when $\mathbf{r}_i = \mathbf{r}_j$, therefore $\overline{\mathbf{M}}_i = V_i\overline{\mathbf{G}}_{ii}$. It was shown that the Fourier spectrum of $\mathbf{E}(\mathbf{r})$ lies on a sphere with radius $m(\mathbf{r})k$, if $m$ is constant in the vicinity of **r**. Therefore at least two sampling points per wavelength in the scatterer are required. The susceptibility is also filtered, either by a mean value filter or a more complicated one, e.g. a Hanning window. This approach is called FCD (Filtered Coupled Dipoles), and a computer code library for evaluation of filtered Green's function is available [45].

Chaumet *et al.* [11] proposed direct Integration of the Green's Tensor (IT) in Eqs. (12), (13). A Weyl expansion of the Green's tensor is performed, transforming it to a form allowing efficient numerical computation of the self-term ($\overline{\mathbf{M}}-\overline{\mathbf{L}}$). They also proposed a correction to the second term in Draine's expression for $C_{abs}$ (Eq. (35)). Extension of their results to a non-isotropic self-term is

$$C_{abs}^{(0)} = 4\pi k\sum_i \left[\mathrm{Im}\left(\mathbf{P}_i \cdot \mathbf{E}_i^{exc*}\right) + \mathrm{Im}\left(\mathbf{P}_i \cdot (\overline{\mathbf{M}}_i - \overline{\mathbf{L}}_i)^* \mathbf{P}_i^*\right)/V_i\right], \tag{58}$$

The corrected second term is based on radiation energy of a *finite* dipole [11]: $\mathrm{Im}(\mathbf{E}_i^{self} \cdot \mathbf{P}_i^*)$, in contrast to a *point* dipole used in the derivation of Eq. (35). One can see that Eqs. (58) and (31) are equivalent. Moreover, both of them are equivalent to Eq. (35) if and only if

$$\overline{\mathbf{M}}_i = \overline{\mathbf{\Lambda}}_i + (2/3)\mathrm{i}k^3 V_i \overline{\mathbf{I}},\ \overline{\mathbf{\Lambda}}_i = \overline{\mathbf{\Lambda}}_i^H. \tag{59}$$

This condition is similar, but not equivalent, to Eq. (36) and is always satisfied for RR, DGF, and LAK. Other polarizability prescriptions satisfy Eq. (59) for real *m*, then both Eqs. (58) and (35) result in zero absorption.

Rahmani, Chaumet, and Bryant [46] proposed a new method (RCB) to determine polarizability based on the known solution of the electrostatic problem for the same scatterer. In the static limit the electric field at any point is linearly related to the incident field

$$\mathbf{E}(\mathbf{r}) = \overline{\mathbf{C}}^{-1}(\mathbf{r})\mathbf{E}^0(\mathbf{r}). \tag{60}$$

Substituting Eq. (60) into Eq. (20) with the static Green's tensor, one can obtain the polarizability, which would give an exact solution in the static limit, as

$$\overline{\alpha}_i^{RCB} = V_i \chi_i \overline{\mathbf{\Lambda}}_i^{-1}, \tag{61}$$

$$\overline{\mathbf{\Lambda}}_i = \overline{\mathbf{C}}_i + \sum_{j\neq i} \overline{\mathbf{G}}^s(\mathbf{r}_i, \mathbf{r}_j) V_j \chi_j \overline{\mathbf{C}}_j^{-1} \overline{\mathbf{C}}_i, \tag{62}$$

where $\overline{\mathbf{C}}_i = \overline{\mathbf{C}}(\mathbf{r}_i)$. This static polarizability then replaces the CM polarizability, and the RR (Eq. (34)) is applied to it [46] to obtain the final polarizability for DDA simulations. It was later shown that RCB polarizabilities differ significantly from CM only for dipoles closer than 2*d* to the interface [47].

In their next manuscript [48] Rahmani *et al.* stated that the previous derivation is correct only if the tensor $\overline{\mathbf{C}}$ is constant inside the particle (e.g. for ellipsoids), since otherwise the polarizability tensor obtained from Eq. (61) is generally not symmetric, which is physically impossible in the static case. This shows that a particle with a non-constant $\overline{\mathbf{C}}$ is not equivalent to any set of physical point dipoles even in the static regime. However, it is equivalent to a set of non-physical dipoles with an asymmetric polarizability. Therefore, the polarization defined by Eq. (61) formally can be used, by itself or with RR, even when $\overline{\mathbf{C}}$ is not constant.

Collinge and Draine [47] empirically combined the RCB prescription with CLDR to get the Surface Corrected LDR (SCLDR):

$$\overline{\boldsymbol{\alpha}}^{SCLDR} = \overline{\boldsymbol{\alpha}}^{RCB}\left(\overline{\mathbf{I}} - \left(\overline{\boldsymbol{\alpha}}^{RCB}/d^3\right)\overline{\mathbf{B}}\right)^{-1}, \tag{63}$$

where $\overline{\mathbf{B}}$ is the correction matrix (analogous to Eq. (50)):

$$B_{\mu\nu} = \delta_{\mu\nu}\left[\left(b_1^{LDR} + b_2^{LDR}m^2 + b_3^{LDR}m^2 a_\mu^2\right)(kd)^2 + (2/3)\mathrm{i}(kd)^3\right]. \tag{64}$$

All methods based on the paper by Rahmani *et al.* [46] are initially limited to very specific shapes of the scatterer (ellipsoids, infinite slabs and cylinders). Expansion of its applicability to other shapes is debatable [48] and would anyway require a preliminary solution of the electrostatic problem for the same shape, which is generally not trivial.

All DDA formulations are schematically depicted in Fig. 1, which also shows interrelations between them. Some formulations can be compared unambiguously in terms of theoretical soundness: one is an improvement of the other, i.e. it employs fewer approximations. Such formulations are depicted in the same column on Fig. 1, while others cannot be compared directly with each other; they give rise to different columns. Comparison between formulations from different columns can and has been made almost exclusively empirically by comparing the accuracy of the simulation results (see Subsection 3.2).

All the above techniques are aimed at reducing discretization errors; only a few aim at reducing shape errors. Some of them employ adaptive discretization (different dipole sizes) to better describe the shape of the scatterer (see Subsection 3.4). Another approach is to average susceptibility in boundary subvolumes. The simplest averaging using the Lorentz-Lorenz mixing rule was proposed by Evans and Stephens [49] for the case of the boundary between the scatterer and its surrounding medium

$$\frac{\chi_i^e}{4\pi\chi_i^e + 3} = f\,\frac{\chi_i}{4\pi\chi_i + 3}, \tag{65}$$

where $\chi_i^e$ is the effective susceptibility, and $f$ is the volume fraction of the subvolume actually occupied by scatterer.

A more advanced averaging, called the Weighted Discretization (WD), was proposed by Piller [13]. It modifies the susceptibility and self-term of the boundary subvolume.[5] The particle surface, crossing the subvolume $V_i$, is assumed linear and divides the subvolume into two parts: the principal $V_i^p$ that contains the center and a secondary $V_i^s$ with susceptibilities $\chi_i^p$, $\chi_i^s$ and electric fields $\mathbf{E}_i^p \equiv \mathbf{E}_i$, $\mathbf{E}_i^s$ respectively. Electric fields are considered constant inside each part and related to each other via a boundary condition tensor $\overline{\mathbf{T}}_i$:

$$\mathbf{E}_i^s = \overline{\mathbf{T}}_i \mathbf{E}_i . \tag{66}$$

---

[5] any subvolume that has non-zero intersection with both the scatterer and the outer medium. All such subvolumes are accounted for.

Then the total polarization of the subvolume can be evaluated as follows:

$$\mathbf{P}_i = \int_{V_i} d^3 r' \chi(\mathbf{r}')\mathbf{E}(\mathbf{r}') = V_i^p \chi_i^p \mathbf{E}_i + V_i^s \chi_i^s \mathbf{E}_i^s = V_i \overline{\chi}_i^e \mathbf{E}_i , \tag{67}$$

$$\overline{\chi}_i^e = \left( V_i^p \chi_i^p \overline{\mathbf{I}} + V_i^s \chi_i^s \overline{\mathbf{T}}_i \right) / V_i . \tag{68}$$

The susceptibility of the boundary subvolume is replaced by an effective one.

The effective self-term is evaluated directly starting from Eq. (3), considering $\chi$ and $\mathbf{E}$ constant inside each part:

$$\overline{\mathbf{M}}_i^{\mathrm{e}} \overline{\boldsymbol{\chi}}_i^{\mathrm{e}} = \int_{V_i^{\mathrm{p}}} \mathrm{d}^3 r' \left( \overline{\mathbf{G}}(\mathbf{r}_i, \mathbf{r}') - \overline{\mathbf{G}}^{\mathrm{s}}(\mathbf{r}_i, \mathbf{r}') \right) \chi_i^{\mathrm{p}} + \int_{V_i^{\mathrm{s}}} \mathrm{d}^3 r' \left( \overline{\mathbf{G}}(\mathbf{r}_i, \mathbf{r}') \chi_i^{\mathrm{s}} \overline{\mathbf{T}}_i - \overline{\mathbf{G}}^{\mathrm{s}}(\mathbf{r}_i, \mathbf{r}') \chi_i^{\mathrm{p}} \right). \tag{69}$$

Piller [13] evaluated the integrals in Eq. (69) numerically. The final equations are the same as Eq. (20), where polarizabilities are obtained from Eq. (18) using effective susceptibilities and self-terms for boundary subvolumes. Hence, WD does not modify the general numerical scheme.

Currently, there are no rigorous theoretical reasons for preferring one formulation over others. However, theoretical analyses of DDA convergence when refining discretization recently conducted by Yurkin *et al.* [32], showed that IT and WD significantly improve the convergence of shape and discretization errors, respectively. Experimental verification of these theoretical conclusions is still to be performed.

### 3.2 Accuracy of DDA simulations

Over the years many results on the accuracy of DDA simulations have been published. It is, however, generally hard to systematically compare the relevant manuscripts because they all use different independent parameters, such as the size parameter $x$, refractive index $m$, or discretization, as a function of which the error is measured. We will describe discretization by the parameter $y = |m| kd$ or $y^{\mathrm{Re}} = \mathrm{Re}(m) kd$. The former is used wherever possible; however, in some cases a description of results is more straightforward in terms of $y^{\mathrm{Re}}$. Accuracy results for scattering by a sphere are summarized in Table 1. All manuscripts on this subject can be divided into two classes: those that fix $x$ and vary $N$ (or equivalently, the number of dipoles per sphere radius $a/d$) with $y$, and those that fix $a/d$ and vary the size parameter with $y$. The former is easier to interpret; the latter is easier to simulate. To facilitate comparison between different methods we provide both $x$ and $a/d$, however one of them is dependent on the other. Some additional information on these results follows below.

Draine and Goodman [3] compared RR, DGF, and LDR for cross sections of a sphere with $a/d = 16$. DGF is generally more accurate than RR. For $|m-1| \leq 1$ LDR gives superior or comparable results to DGF, for $m = 2 + \mathrm{i}$ LDR and DGF are comparable, and for $m = 4 + 3\mathrm{i}$ DGF is preferable over LDR. In the review of LDR DDA, Draine and Flatau [4] summarized that for $|m| \leq 2$ cross sections can be evaluated to accuracies of a few percent provided $y \leq 1$. In that case differential cross sections have satisfactory accuracy: relative errors up to 20-30%, but only where the absolute value of the differential cross sections is small. For spheres, such results are obtained even for $|m| \leq 3$. Comparison of CLDR to LDR [43] only results in minor differences. Generally CLDR results in slightly better accuracy for $C_{sca}$, but worse for $C_{abs}$.

Piller and Martin [44] compared FCD to LAK by studying the dependence of the mean relative error of the far-field electric fields ($\Psi$) on $y$ for spheres with $x = \pi$, $2\pi$ and $m = 1.5$. It was shown that FCD (with a Hanning window filter for the electric permittivity $\varepsilon$) is roughly 3 times more accurate than LAK in the range $0.7 \leq y \leq 2.5$ and gives similar accuracies for $y \leq 0.4$ (for larger spheres). Comparison of WD to traditional methods [13] was performed for spheres with $x = \pi$, $2\pi$ and $m = 1.32$, $2.1 + 0.7\mathrm{i}$. LAK was used to determine polarizabilities. For $m = 1.32$ in the range $0.4 \leq y \leq 1.3$ overall accuracy was only slightly improved, but error peaks for certain values of $y$ were smoothed out. For $m = 2.1 + 0.7\mathrm{i}$ accuracy was improved 4-5 times over the whole range $y \leq 1.3$. Piller also showed [10] that a combination of WD and FCD gives even better results. Generally FCD decreases the negative effects of $\mathrm{Re}(\varepsilon)$ on accuracy and WD those of $\mathrm{Im}(\varepsilon)$.

Rahmani *et al.* [48] showed that RCB was clearly superior to CM in calculating cross sections for fixed $a/d = 16$ and $m$ from $1.8 + 0.4\mathrm{i}$ to $7.4 + 9.4\mathrm{i}$ in the range $y \leq 1$. Two corrections (LDR and RR) over the static case were compared, and they gave similar overall results. Improvement of overall accuracy compared with CM was 2-5 times in all cases studied. For a thin slab, it was shown [46,48] that the internal fields calculated using RCB differ from those by CM mostly near the interfaces, where RCB yields much smaller errors, almost the same as far from interfaces.

Collinge and Draine [47] compared LDR, RCB, and SCLDR in calculations of cross sections of spheres with $a/d = 12$. It was shown that for $m = 1.33 + 0.01\mathrm{i}$, LDR and SCLDR are superior in the range $y \leq 0.8$, while for $m = 5 + 4\mathrm{i}$, SCLDR and RCB are superior.

Convergence of cross sections for spheres and ellipsoids for increasing $N$ with fixed $x$ and different $m$ (from $1.33+0.01\mathrm{i}$ to $5+4\mathrm{i}$) also was studied. SCLDR showed the most stable results for all cases, being the most or close to the most accurate one; however, for ellipsoids with large Im($m$) RCB gave significantly more accurate results for $C_{sca}$, especially for larger $y$.

Performance of the DDA for more complex shapes also was studied by different authors. Flatau *et al.* [50] compared DDA simulations for a bisphere with an exact solution from a multipole expansion. For $m=1.33+0.01\mathrm{i}$, $a/d=16$, and $y\le 0.8$, LDR was several times more accurate than DGF and resulted in errors of less than 0.5% for both $C_{sca}$ and $C_{abs}$. Xu and Gustafson [51] made a similar but much more extended study of LDR. For $m=1.6+0.008\mathrm{i}$, $a/d=25$, and $y\le 0.4$, errors in $C_{ext}$, $C_{abs}$, and $\langle\cos\theta\rangle$ are within 10%. For $y=0.81$, errors in the angular dependence of $S_{11}$ are up to 20% while $S_{12}$ and $S_{21}$ were completely wrong. For $m=2.5+0.02\mathrm{i}$, errors in cross sections exceed 10% for $y\ge 0.3$. Errors in the angular dependencies of the Mueller matrix elements are within 10-20% for $y=0.3$ and increase rapidly with increasing $y$. For a fixed $x=3$ and $m=1.6+0.004\mathrm{i}$, errors in $C_{ext}$, $C_{abs}$, and $\langle\cos\theta\rangle$ decrease from 10% to 1% while $y$ decrease from 1 to 0.2. For $y=0.33$, the angular dependence of $S_{11}$ is in good agreement with the rigorous solution, while $S_{12}$ and $S_{21}$ differ significantly for certain orientations of the bisphere.

Hage and Greenberg [14] compared LAK to experimental results obtained from microwave experiments on porous cubes. Using $m=1.362+0.005\mathrm{i}$, $y=0.64$ and $N=5504$, they obtained a difference of less than 40% with the experimental results of angular scattering patterns, except for deep minima. Light scattering of cubes, tiles, and cylinders with similar parameters also was studied and comparable differences between experiment and theory were obtained. Theoretical errors were estimated to be less than 10%, except for deep minima.

Iskander *et al.* [34] conducted a limited test of LAK for small elongated spheroids, comparing the results to those obtained using an iterative extended boundary condition method. Using $N=64$, calculations were performed for aspect ratios up to 20 with maximum size parameter of the long axis being 10 and 0.5 for $m=1.33+0.01\mathrm{i}$ and $1.76+0.28\mathrm{i}$ respectively. Errors in scattering cross section were 21% and 11%, respectively. Ku [52] compared LAK with CM and the $a_1$-term for different shapes, but his conclusions are based on a large parameter $y$ (up to 2), and are therefore suspicious and not further discussed here.

Andersen *et al.* [53] studied the performance of the DDA for Rayleigh-sized clusters of a few spheres (most DDA formulations are then equivalent to CM). Several constituent materials were tested, all with high refractive indices in the studied region. It was shown that the DDA failed to converge using the fixed computational resources for very high (up to 13.0) and very low (down to 0.12) Re(*m*); up to 30 dipoles were used per diameter of a single sphere.

It can be concluded that particles with more complex shapes than spheres are more difficult to model with the DDA, leading to larger errors for the same *m* and *y*. This effect can be explained in general by the increase of surface to volume ratio and hence larger fraction of boundary subvolumes [32]. Another possible reason is complex regions, e.g. contact between two particles in a cluster, where rapid variation of the electric field deteriorates the overall accuracy. There is, however, a notable exception from this general tendency. Shapes, which can be modeled exactly by a set of cubical dipoles, e.g. a cube, can be simulated using the DDA much more accurately than spheres, especially for small *y* [32].

Draine and Flatau [4] have introduced a "rule of thumb" for discretization: use 10 dipoles per wavelength in the medium (i.e. either *y* or $y^{\mathrm{Re}}$ equal to 0.63, depending on the interpretation). Though it is widely used, the accuracy of the results, when using such discretization, is hard to deduce *a priori*. Draine and Flatau themselves derived an estimate of the error based on a set of test simulations. This estimate is described above and mentioned in Table 1; it is usually cited as a "few percent accuracy in cross sections." However, it may significantly over- or under-estimate the error, especially for large size parameters. Moreover, it does not completely account for the dependence on *m,* even in the stated range of its application ($|m| \leq 2$), since DDA accuracy deteriorates rapidly with increasing *m* (see Table 1). Still, the rule of thumb is good first guess for many applications.

Most studies of DDA accuracy are limited to integral scattering quantities and, at most, the angular dependence of $S_{11}$. In only a few manuscripts are other scattering quantities studied. For instance, Singham [54] simulated the angular dependence of Mueller matrix element $S_{34}$ for spheres and less compact particles, using CM polarizability. It was shown that an accurate simulation of this element requires smaller values of *y* than for $S_{11}$. For $x = 1.55$ and $m = 1.33$ a calculation of $S_{11}$ was accurate already for $y = 0.8$, while $y \leq 0.2$ was required for $S_{34}$. It was also reported that for less compact objects like discs and rods, the required *y* was larger, 0.4 and 0.55 respectively, because of the smaller interaction between the dipoles. However, Hoekstra and Sloot argued [55] that this effect is mostly caused by the pronounced $S_{34}$ sensitivity to surface roughness, which is significant for smaller size if *y* is fixed. They showed that for $x = 10.7$ and $m = 1.05$, very high accuracy is achieved with $y = 0.66$ because of the larger number of dipoles used.

Internal fields are an intermediate result in the DDA. They cannot be directly compared to the experimental results; however, all measured scattering quantities are derived from them. Therefore, a study of their accuracy can reveal greater understanding of the nature of DDA errors. Hoekstra *et al.* [56] performed such a study for LAK polarizability. Three spheres were examined with $x = 9$, 9, 5 and $m = 1.05$, $1.33 + 0.01\mathrm{i}$, $2.5 + 1.4\mathrm{i}$ respectively. Values of $y$ were 0.44, 0.42, and 0.51 respectively. The most significant errors in the amplitude of the internal field were localized at the boundary of the spheres with maximum relative errors of 3.4%, 19%, and 120% respectively. Errors in $S_{12}$, $S_{33}$, $S_{34}$ were significant only for the third sphere. It was shown that for a given $y^{\mathrm{Re}}$ these errors rapidly increase with $m$ but only slightly depend upon $x$ in the range from 1 to 10. Moreover, the DDA is capable of reproducing resonances of Mie theory, although their positions are slightly shifted (less than 1% in $m$).

Druger and Bronk [57] studied the accuracy of the internal fields for single and coated spheres. They used $x = 1.5$, $m \le 1.8$, and CM polarizability. Errors in the internal fields were localized at the interfaces, with average errors larger than 30% for a single sphere with $m = 1.8$ and $y = 0.17$, and less than 7% for a single and concentric sphere with $m = 1.3$ and $y = 0.08$. The core of the concentric sphere has $m = 1.1$ and its diameter is half the total diameter. The angular dependence of the absolute values of $S_1$ and $S_2$ had significant errors in the side- and backscattering. It can be concluded that shape errors contribute mostly to the internal fields near the boundary, and increase with $m$.

All the literature discussing DDA accuracy shows errors as a function of input parameters and discretization, which is the most straightforward way. The only exception so far is the rule of thumb, which is too general and approximate to be applied in many particular cases. A more useful way to present errors is to fix the desired accuracy for certain input parameters and find the discretization that results in such accuracy. Such an analysis can be applied directly to practical calculations and can be used to derive rigorous estimates of DDA computational requirements [58].

In a number of manuscripts the origin of errors in the DDA was examined to try to separate and compare shape and discretization errors [49,59-62]; however, no definite conclusions were reached. The uncertainty was due to the indirect methods used that have inherent interpretation problems. Recently, Yurkin *et al.* [63] proposed a direct method to separate shape and discretization errors, which can be used to study their fundamental properties. This method also can be applied to study the performance of different formulations aimed at decreasing shape errors, e.g. WD. For example, it has been shown that the maximum errors of $S_{11}(\theta)$ for a sphere with $x = 5$ and $m = 1.5$, discretized using 16 dipoles per diameter ($y = 0.93$), are mostly due to shape errors. However the same is not true for all measured quantities. In another manuscript [32] it was suggested that the discretization error should decrease more rapidly with decreasing *y* than shape errors. However, it is still hard to deduce *a priori* the importance of shape errors for a certain scatterer and *y*; hence, further systematic quantitative study is required.

### 3.3 The DDA for clusters of spheres

There are two main peculiarities when the DDA is applied to clusters of spheres. First, such particles are generally less compact, yielding smaller interactions between dipoles. This leads to a smaller condition number of the DDA interaction matrix and hence faster convergence of the iterative solver (see Section 4.1). Second, when the constituent spheres are small compared to the wavelength, each sphere can be modeled as one spherical subvolume, yielding some theoretical simplifications.

A general theory exists [64] based on the Mie theory (Generalized Multiparticle Mie solution (GMM) [65]) that allows for highly accurate simulations of clusters of spheres. However, when many small spheres are used one wants to minimize the number of unknowns in the linear system. Direct reduction of the GMM to the lowest order (using only the first order expansion coefficients) leads to DDA + CM [64]. Improving accuracy in the GMM is done by accounting for higher multipole moments, while the DDA introduces higher order corrections to the coefficients of the linear system. It is not clear how the accuracy of these two methods compare with each other; however, the former should lead to a formulation similar to a coupled multipole method (Subsection 3.4) with a larger number of unknowns. DDA-based methods (starting usually with the integral equations introduced in Section 2) should be successful in making the formulation more accurate without increasing the number of unknowns, which is the goal for *large* clusters of *small* spheres. Moreover, the DDA may employ fast algorithms for solving the linear system. In this setting, the Fast Multipole Method (FMM) (see Subsection 4.5) seems most promising.

It should be noted, however, that a cluster having a small size parameter (i.e. in the electrostatic approximation) does not imply that all expansion coefficients, except the first one, are negligible. This is because the size of the constituent particles is also very small and the fields inside them are far from constant, especially when the spheres are located close to each other and have large refractive indices [66]. Therefore, the DDA does have some principal difficulties of calculating scattering by clusters of spheres. Mackowski [67], for instance, found that for some systems composed of spheres much smaller than the wavelength, up to 10 expansion terms were necessary to achieve convergence. In studies of osculating spheres, Ngo *et al.* [68] proved that the GMM could be chaotic and were able to calculate Lyapunov exponents, and that the slow convergence for the touching spheres was the result of the system lying in an attractor region. A recent paper by Markel *et al.* [69]

presented computationally efficient modifications of the GMM in the static limit and demonstrated the insufficiency of the DDA to compute scattering properties of fractal aggregates accurately. However, Kim *et al.* [70] showed that the DDA is satisfactory in calculating the static polarizability of dielectric nanoclusters, especially of clusters with a large number of constituents.

The development of DDA-based methods for calculating light scattering by clusters of small spheres was started by Jones [71,72], who developed a method similar to CM. Iskander *et al.* [34] used a method equivalent to LAK to calculate scattering of chained aerosol clusters. This subject was further investigated by Kosaza [73,74]. Lou and Charalampopoulos [75] (LC) further improved the calculations of the interaction term and scattering quantities. Starting from an integral equation for the internal field equivalent to Eq. (1), they assumed Eq. (11). After that the integrals in Eqs. (12) and (13) over spherical subvolumes can be evaluated analytically. The result for the interaction term is the following:

$$\overline{\mathbf{G}}_{ij}^{(0)} = \eta(ka)\overline{\mathbf{G}}(\mathbf{r}_i,\mathbf{r}_j)\,, \tag{70}$$

where a correction function $\eta$ is defined as

$$\eta(x) = 3\frac{\sin x - x\cos x}{x^3} = 1-(1/10)\,x^2 + \mathrm{O}(x^4)\,. \tag{71}$$

Eq. (30) also is evaluated analytically, yielding

$$\mathbf{F}^{(0)}(\mathbf{n}) = -\mathrm{i}k^3\eta(ka)(\overline{\mathbf{I}} - \hat{n}\hat{n})\sum_i \mathbf{P}_i \exp(-\mathrm{i}k\mathbf{r}_i \cdot \mathbf{n})\,, \tag{72}$$

$$C_{ext}^{(0)} = 4\pi k\,\eta(ka)\sum_i \mathrm{Im}\left(\mathbf{P}_i \cdot \mathbf{E}_i^{inc*}\right). \tag{73}$$

The following expression for $C_{abs}$ is stated without derivation:

$$C_{abs}^{(0)} = 4\pi k\,\eta(ka)\sum_i \mathrm{Im}(\mathbf{P}_i\mathbf{E}_i^*)\,. \tag{74}$$

Markel *et al.* [76] applied the DDA to fractal clusters of spheres, and studied their optical properties. However, they have not fixed the polarizability of a single dipole but rather treated it as a variable, calculating the dependence of a cluster's optical characteristics upon it. Pustovit *et al.* [77] argued that the DDA is inaccurate for touching spheres. They developed a hybrid of the DDA and the GMM, which considers only pair interactions between spheres (as the DDA) but, when calculating them, accounts for higher multipole terms. This formulation can be considered as the one providing a more accurate evaluation of the interaction term (Eq. (13)), and hence similar to LC.

LC was compared to DGF and LAK in a $C_{sca}$ computation of a cluster of 10 particles for $m = 1.7 + 0.7\mathrm{i}$ and $0.05 \le ka \le 0.5$. Differences between DGF and LAK are less than 1% (as expected), while the difference between LC and LAK increases quadratically with $ka$, reaching 10% for $ka = 0.5$. However, as no exact (e.g. GMM) solution is presented, the accuracy of each individual method is not clear.

Okamoto [42] tested the $a_1$-term method for clusters of up to 3 touching spheres. No effective medium is needed in this case, making the method sounder. It was shown that the $a_1$-term is clearly superior to LDR in cross-sections calculations, when each sphere is treated as a single dipole. Errors of the $a_1$-term are less than 10% for $y \leq 1.2$ when $m = 1.33 + 0.01\mathrm{i}$. For three collinear touching spheres the errors are 30% and 40% for $y \leq 1.9$ and 2.8 when $m = 1.33 + 0.01\mathrm{i}$ and $2 + \mathrm{i}$ respectively. However, errors do not seem to diminish significantly for small $y$ (results are presented only down to $y = 0.2$). Therefore, the $a_1$-term seems suitable for obtaining quick crude estimations of cross sections.

In the sequel of this subsection we mention several applications of the DDA to scattering from clusters of spheres. It was applied to describe the scattering by astrophysical dust aggregates [78,79] using the $a_1$-term method. Hull *et al.* [80] applied CM DDA to Diesel soot particles. LC was applied [81] to the computation of light scattering by randomly branched chain aggregates. Lumme and Rahola [40] studied scattering properties of clusters of large spheres (each modeled by a set of dipoles) with the $a_1$-term method considering astrophysical applications. Hage and Greenberg [35] studied scattering by porous particles, which were modeled as clusters of *cubical* cells making their method equivalent to standard LAK. Recently the DDA with LDR was used [82] to model scattering by porous dust grains and compare them to approximate theories, e.g. effective medium theories. It also was used to study light scattering by fractal aggregates [83], especially its dependence on the internal structure [84].

### 3.4 Modifications and extensions of the DDA

Bourrely *et al.* [85] proposed to use small $d$ to minimize surface roughness, but larger dipoles inside the particle. Starting with small dipoles with CM polarizability, one dipole is combined with 6 adjacent ones (if they all have the same polarizability) producing a dipole, located at the same point but with a 7 times larger polarizability. This operation is repeated while possible. Interaction terms are considered in their simplest form (Eq. (14)). This method allows the decrease of the shape errors with only a minor increase in the number of dipoles. The authors showed that this method is more than two times more accurate than CM for some test cases.

Rouleau and Martin [86] proposed a generalized semi-analytical method. A dynamic grid is used to evaluate the integral in Eq. (1). First, a static grid is built inside the particle. Then each point on the static grid is used as an origin of a spherical coordinate system, and the particle is approximated by an ensemble of volume elements in these spherical coordinates. As usual, the polarization inside each subvolume is assumed constant, but Eq. (13) can be evaluated analytically in spherical coordinates. Polarization inside a subvolume is obtained by interpolation of its values at the points of the static grid. In addition, adaptive gridding is employed, where smaller subvolumes are used at the boundary of the particle.

Mulholland *et al.* [87] proposed a Coupled Electric and Magnetic Dipole method (CEMD), where a magnetic dipole is considered at each subvolume together with an electric dipole. Polarizabilities are derived from the $a_1$ and $b_1$ terms of the Mie theory. CEMD requires two times more variables in the linear system, since the electric and magnetic fields are interconnected. Lemaire [88] went further and developed the coupled multipole method, considering also the electric quadrupole. Addition of the electric quadrupole can be considered as a more accurate evaluation of the interaction term in Eq. (13), as compared to Eq. (14). It results in even better accuracy than CEMD, but at the expense of additional computation time. The major disadvantage of all these four methods is that the matrix of the system of linear equations does not seem to have any special form, suitable for faster algorithms (see Section 4). Therefore computational costs are much larger compared to regular methods, thus limiting their practical use. In what follows, several DDA extensions are mentioned without further discussion.

The theoretical basis for application of the DDA to optically anisotropic particles was summarized by Lakhtakia [89]. Loiko and Molochko [90] applied the DDA to study light scattering by liquid-crystal spherical droplets. Smith and Stokes [91] used the DDA to calculate the Faraday effect for nanoparticles. Researchers in the electrical engineering community applied MoM (in a variation that is equivalent to the DDA) to anisotropic scatterers [92,93].

Rectangular parallelepipeds can be used as subvolumes in the DDA [11,23,43]. This allows an accurate description of light scattering by particles with large aspect ratios, using fewer dipoles and is also compatible with FFT techniques (Subsection 4.4).

Khlebtsov [94] proposed a simplification of the DDA, based on the assumption that all polarizations are parallel to the incident electric field. The number of variables is thus reduced three times, however at a cost of accuracy. Moreover, depolarization is completely ignored.

Markel [95] analytically solved the DDA equations for scattering by an infinite one-dimensional periodic dipole array. This approach is similar to the one used in obtaining the LDR formulation for dipole polarizability [3].

Chaumet *et al.* [96] generalized the DDA to periodic structures, and further to defects in a periodic grating on a surface [97]. The idea of using the complex Green's tensor in the standard DDA formulation was summarized by Martin [98].

Yang *et al.* [99] used the DDA to calculate surface electromagnetic fields and determine Raman intensities for small metal particles of arbitrary shape.

Lemaire and Bassrei [100] showed that the shape of an object can be reconstructed from the measured angle dependence of scattered intensities. This procedure can be thought of as an inversion of the dependence between dipole polarizabilities and scattering. This dependence is taken from the DDA. A similar idea is used in recent manuscripts on optical tomography [101-103].

Zubko *et al.* [104] modified the Green's tensor used in the DDA to study the backscattering of debris particles. They showed that the far-field part of the Green's tensor is responsible for both the backscattering brightness surge and the negative polarization branch.

# 4 Numerical considerations

In this Section the numerical aspects of the DDA are discussed. One should keep in mind, however, that final simulation times depend not only on the chosen numerical methods but also on the particular implementation. Recently, Penttila *et al.* [105] have compared four different computer programs for the DDA. These are based on almost identical numerical methods: the Krylov-subspace iterative method (Section 4.1) combined with a FFT acceleration of the matrix-vector product (Section 4.4). However, simulation times may differ by several factors. Optimizations of computer codes are not further discussed in this review.

## 4.1 Direct *vs.* iterative methods

There are two general types of methods to solve linear systems of equations $\mathbf{A}\mathbf{x} = \mathbf{y}$, where **x** is an unknown vector and **A** and **y** are known matrix and vector, respectively: direct and iterative [106]. Direct methods give results in a fixed number of steps, while the number of iterations required in iterative methods is generally not known *a priori*. The most usual example of a direct method is *LU* decomposition, which allows quick solving for multiple **y** once the decomposition is performed. Iterative methods are usually faster, less memory consuming and numerically more stable. However, iterative methods cannot be considered superior over direct, since they strongly depend on the problem to solve [107].

For a general $n \times n$ matrix (in DDA $n = 3N$) computation time of *LU* decomposition is $O(n^3)$ and storage requirements $O(n^2)$, while computation time for one iteration is $O(n^2)$ [107]. Iterative methods for a general matrix converge in $O(n)$ iterations, although some of them may not converge at all. However, in many cases satisfactory accuracy can be obtained after a much smaller number of iterations. In these cases, iterative methods can provide significant increases in speed, especially for large $n$. Most iterative methods access the matrix $\mathbf{A}$ only through matrix-vector multiplication (sometimes also with the transposed matrix), which allows the construction of special routines for calculation of these products. Such routines may decrease memory requirements, since it is no longer necessary to store the entire matrix, especially for matrices of special form (see Subsection 4.3). A special structure of the matrix may also allow acceleration of the matrix-vector product from $O(n^2)$ to $O(n\ln n)$ (see Subsections 4.4, 4.5). However, the same applies to direct methods (see Subsection 4.3).

Throughout DDA history, mostly iterative methods were employed (however see Subsection 4.6). At first, they were used to accelerate computations [1], but they also allowed larger numbers of dipoles to be simulated [6,108], since storage of the entire matrix is prohibitive for direct methods. The most widely used iterative methods in the DDA are Krylov-space methods, such as [107] Conjugate Gradient (CG), CG applied to the Normalized equation with minimization of Residual norm (CGNR), Bi-CG, Bi-CG STABilized (Bi-CGSTAB), CG Squared (CGS), Generalized Minimal RESidual (GMRES), Quasi-Minimal Residual (QMR), Transpose Free QMR (TFQMR), and Generalized Product-type methods based on Bi-CG (GPBi-CG) [109].

An important part of the iterative solver is preconditioning, which effectively decreases the condition number of the matrix $\mathbf{A}$ and therefore speeds up convergence. However, this requires additional computational time during both initialization and each iteration. Preconditioning of the initial system can be summarized as [107]

$$\mathbf{M}_1\mathbf{A}\mathbf{M}_2^{-1}(\mathbf{M}_2\mathbf{x}) = \mathbf{M}_1\mathbf{y}\,, \tag{75}$$

where $\mathbf{M}_1$ and $\mathbf{M}_2$ are left and right preconditioners, respectively. Preconditioners should either allow fast inversion or be integrated into the iteration process. The simplest preconditioner of the first type is the Jacobi (point), which is just the diagonal part of matrix $\mathbf{A}$. An example of the second type of preconditioner is the Neumann polynomial preconditioner of order $l$:

$$\mathbf{M} = \sum_{j=1}^{l}(\mathbf{I}-\mathbf{A})^j\,. \tag{76}$$

QMR and Bi-CG can be made to employ the Complex Symmetric (CS) property of the DDA interaction matrix to halve the number of matrix-vector multiplications [110] (and thus computational time). Lumme and Rahola [40] were the first to apply QMR(CS) to the DDA and compared it with CGNR. They used $m$ from $1.6+0.1\mathrm{i}$ to $3.0+4.0\mathrm{i}$, and $x$ from 1.3 to 13.5, corresponding to *N* from 136 to 20336. For all cases studied QMR(CS) was 2-4 times faster than CGNR.

Rahola [9] further studied QMR(CS) and compared it to CGNR, Bi-CG(CS), Bi-CGSTAB, CGS, GMRES (full and with different memory length). For a "typical small problem" (parameters were not specified, unfortunately) the convergence of different methods was tested and QMR(CS) along with Bi-CG(CS) showed the best results. Although full GMRES was able to converge in fewer iterations, GMRES with as much as 40 memory lengths was slower than QMR(CS).

Flatau [111] reviewed the use of iterative algorithms in the DDA and tested many of them, together with several preconditioners. He calculated scattering of a homogenous sphere with $x = 0.1$ and *m* from 1.33 up to $5 + 0.0001\mathrm{i}$, $x = 1$ and *m* from 1.33 up to $1.33 + \mathrm{i}$ and $3 + 0.0001\mathrm{i}$. Left (L) and right (R) Jacobi-, and first-order Neumann polynomial preconditioners were tested. Unfortunately the number of dipoles *N* was not specified, which hampers comparison with other studies. For small particles CG(L) was superior for all refractive indices studied. CG and CG(R) showed similar results, while CGNR(L) and Bi-CGSTAB(L) were about 4 times slower. For $x = 1$ Bi-CGSTAB(L) was superior while Bi-CGSTAB,(R) and CGS,(L),(R) were slightly worse. TFQMR (both with and without Jacobi preconditioner) was 3-4 times slower. The first-order Neumann preconditioner showed unsatisfactory results. It was concluded that Bi-CGSTAB(L) is the most satisfactory choice for the DDA, and that method is the default one used in the DDSCAT program [6].

Recently Fan *et al.* [112] have compared GMRES, QMR(CS), Bi-CGSTAB, GPBi-CG, and Bi-CG(CS). They tested them on wavelength-sized scatterers (*x* up to 10) with *m* up to $4.5 + 0.2\mathrm{i}$, and concluded that GMRES with memory depth 30 was the fastest, although it required four times more memory than the other methods. However, only the times of the matrix-vector product was compared, while other parts of the iteration may also take significant time, especially for GMRES(30). Choosing from less memory-consuming methods, QMR(CS) and Bi-CG(CS) showed a better convergence rate than Bi-CGSTAB and GPBi-CG, especially when $|m| > 2$. Moreover, the authors pointed out some flaws in the comparison by Flatau [111], making his conclusions insufficient.

Yurkin *et al.* [113] employed QMR(CS), Bi-CG(CS), and Bi-CGSTAB to simulate light scattering by spheres with *x* up to 160 and 40 for $m = 1.05$ and 2, respectively. It was shown that convergence of the iterative methods becomes very slow with increasing *x* and *m* (up to $10^5$ iterations are required), and none of them is clearly preferable to the others. Moreover, there seems to be no systematic dependency of the choice of the best iterative solver on $x$ and $m$; however, the difference in computational time was less than a factor of two, except for the largest $x$ and $m$ studied.

Rahola [114] showed that the spectrum of the integral scattering operator for any homogenous scatterer is a line in the complex plane going from 1 to $m^2$, except for a small amount of points, which corresponds to refractive indices that cause resonances for the specific shape. The spectrum of **A** is similar, since this matrix is obtained in the DDA by discretization of the integral operator (see also [9]). Assuming that the spectrum of **A** exactly lies on the specified line, it was shown that an estimate for the optimal reduction factor[6] $\gamma$ can be given as

$$\gamma = \frac{1}{\left|1 + \sqrt{\frac{2}{m^2-1}} + \frac{2}{m^2-1}\right|}. \tag{77}$$

Eq. (77) is an approximation valid for small particle sizes, where no, or only few, resonances are present. However, in all cases the spectrum of **A** resembles the spectrum of the linear operator, which is defined by shape, size and refractive index of the scatterer. Therefore, the spectrum, and thus convergence, should not depend significantly on the discretization. This fact was confirmed empirically in other manuscripts [9,63].

Budko and Samokhin [115] generalized Rahola's results to arbitrary inhomogeneous and anisotropic scatterers. They described a region in the complex plane that contains the whole spectrum of the integral scattering operator. This region depends only on the values of $m$ inside the scatterer and does not depend on $x$. They showed that for purely real $m$ or for $m$ with very small imaginary part this region may come close to the origin, therefore the spectrum may contain very small eigenvalues for particles larger than the wavelength. This may explain the extremely slow convergence of the iterative solver for real $m$ and large $x$, which was recently obtained in numerical simulations [113]. Based on the analysis of the spectrum of the integral scattering operator for particles much smaller than the wavelength, Budko *et al.* [116] proposed an efficient iteration method for this particular case.

It can be concluded that there are several modern iterative methods (QMR(CS), Bi-CG(CS), and Bi-CGSTAB) that have proved to be efficient when applied to the DDA. However, none of them can be claimed superior to the others, and one should test them for particular light-scattering problems. Moreover, except for the simplest cases, preconditioning of the DDA interaction matrix is almost not studied, while there is a need for it for large $x$ and $m$, since then all methods converge extremely slowly or even diverge. It seems to us that the next major numerical advance in the DDA will be achieved by developing a powerful preconditioner for the DDA matrix.

---

[6] Norm of the residual is decreased by this factor every iteration.

A large number of dipoles requires large computational power and, hence, parallel computers are commonly used, e.g. [108,113]. Parallel efficiency is not discussed here, but for iterative solvers, it is generally close to 1 [117]. However, this is not true for all preconditioners [107], and hence heavy preconditioners requiring large computational time in combination with a parallel DDA implementation should be employed with caution.

### 4.2 Scattering order formulation

The Rayleigh-Gans-Debye (RGD) approximation [27] consists in considering $\mathbf{E}(\mathbf{r})$ equal to $\mathbf{E}^{inc}(\mathbf{r})$. $\mathbf{F}(\mathbf{n})$ is then obtained directly from Eq. (24). Generalization of the RGD approach is obtained by iteratively solving the integral equation (1), which can be rewritten as

$$\mathbf{E}(\mathbf{r}) = \mathbf{E}^{inc}(\mathbf{r}) + \mathbf{\Lambda}\mathbf{E}(\mathbf{r}), \tag{78}$$

where $\mathbf{\Lambda}$ is a linear integral operator describing the scatterer. The iterative scheme is readily obtained by inserting the current ($l$-th) iteration of the electric field $\mathbf{E}^{(l)}(\mathbf{r})$ into the right side of Eq. (78) and calculating the next iteration in the left side:

$$\mathbf{E}^{(l+1)}(\mathbf{r}) = \mathbf{E}^{inc}(\mathbf{r}) + \mathbf{\Lambda}\mathbf{E}^{(l)}(\mathbf{r}). \tag{79}$$

The starting value is taken the same as in RGD, $\mathbf{E}^{(0)}(\mathbf{r}) = \mathbf{E}^{inc}(\mathbf{r})$, and the general formula for the solution is the following:

$$\mathbf{E}(\mathbf{r}) = \sum_{l=0}^{\infty} \mathbf{\Lambda}^{l}\mathbf{E}^{inc}(\mathbf{r}), \tag{80}$$

which is a direct implementation of the well-known Neumann series:

$$(\mathbf{I} - \mathbf{\Lambda})^{-1} = \sum_{l=0}^{\infty} \mathbf{\Lambda}^{l}, \tag{81}$$

where $\mathbf{I}$ is the unitary operator. A necessary and sufficient condition for Neumann-series convergence is

$$\|\mathbf{\Lambda}\| < 1. \tag{82}$$

Physical sense of this iterative method lies in successive calculations of interaction between different parts of the scatterer. The zeroth approximation (or RGD) accounts for no interaction; the first approximation considers the influence of scattering of each dipole on the others once, and so on. Eq. (82) states that the interaction inside the scatterer should be small, but not as small as required for the applicability of RGD ($\|\mathbf{\Lambda}\| \ll 1$). In scattering problems, especially in quantum physics, Eq. (80) is called the Born expansion.

Although theoretically clear, the Born expansion is not directly applicable [118], since each successive iteration requires analytical evaluation of multidimensional integrals with rising complexity, which quickly becomes unfeasible even for the simplest scatterers. The latest result is probably that of Acquista [118], who evaluated the Born expansion for a homogenous sphere up to second order. Therefore, realistic application of the Born expansion does require discretization of the integral operator, which is naturally done in the DDA.

A Scattering Order Formulation (SOF) of the DDA was developed independently by Chiappetta [119] and Singham and Bohren [12,120] by applying the Neumann series to Eq. (17). $\mathbf{\Lambda}$ is then a matrix defined as $\mathbf{\Lambda}_{ij} = \overline{\mathbf{G}}_{ij}\overline{\boldsymbol{\alpha}}_j$, where each element is a dyadic, which can be expressed as a 3×3 matrix. An explicit check of Eq. (82) for a certain scatterer is not feasible numerically, however de Hoop [121] derived a sufficient condition for scalar waves:

$$2\pi(kR_0)^2 \max_{\mathbf{r}} \left|\chi(\mathbf{r})\right| < 1, \tag{83}$$

where $R_0$ is the radius of the smallest sphere circumscribing the scatterer. Although not directly applicable to light scattering, Eq. (83) can be used as an estimate.

The range of size parameter and refractive index where SOF converges is limited [120]. Moreover, even when SOF converges, more advanced iterative methods converge faster (see Subsection 4.1). However, SOF has clear physical sense and can be used to study the importance of multiple scattering.

### 4.3 Block-Toeplitz

A square matrix $\mathbf{A}$ is called Toeplitz if $A_{ij} = a_{i-j}$, i.e. matrix elements on any line parallel to the main diagonal are the same [106]. In a Block-Toeplitz (BT) matrix (of order $K$) elements $a_i$ are not numbers, but square matrices themselves:

$$\mathbf{A} = \begin{bmatrix} \mathbf{a}_0 & \mathbf{a}_1 & \cdots & \mathbf{a}_{K-1} \\ \mathbf{a}_{-1} & \mathbf{a}_0 & \ddots & \vdots \\ \vdots & \ddots & \ddots & \mathbf{a}_1 \\ \mathbf{a}_{-K+1} & \cdots & \mathbf{a}_{-1} & \mathbf{a}_0 \end{bmatrix}. \tag{84}$$

A 2-level BT matrix has BT matrices as components $\mathbf{a}_i$. Proceeding recursively a Multilevel BT (MBT) matrix for any number of levels is defined.

Let us consider a rectangular lattice $n_x \times n_y \times n_z$, numbered in the following way

$$i = n_y n_z (n_x - 1) i_x + n_z (n_y - 1) i_y + n_z i_z, \tag{85}$$

where $i_\mu \in \{1,\ldots,n_\mu\}$ indicates the position of the element along the axes. Let us also define the vector index $\mathbf{i} = (i_x, i_y, i_z)$. Then one can verify that the interaction matrix in Eq. (20), defined by Eq. (13), satisfies the following:

$$\overline{\mathbf{G}}_{ij} = \overline{\mathbf{G}}_{ji} = \overline{\mathbf{G}}'_{\mathbf{i}-\mathbf{j}} . \qquad (86)$$

This equation alone can be used to greatly reduce the storage requirements of iterative methods by use of indirect addressing. Further improvement is to note that Eq. (86) defines a symmetric 3-level BT matrix (orders of subsequent levels – $n_x$, $n_y$, $n_z$) whose smallest blocks are 3×3 matrices (dyadics) $\overline{\mathbf{G}}_{ij}$.

A rectangular lattice is not much of a restriction, since any scatterer can be embedded in an appropriate rectangular grid. However, additional "empty" dipoles should be introduced to build up the grid up to the full parallelepiped. Moreover, position and size of the dipoles cannot be chosen arbitrarily to better describe the shape of the scatterer. This is especially problematic for highly porous particles or clusters of particles, where the monomer has a size comparable to a single dipole. For all other cases these restrictions are minor compared to the large increase in computational speed, imposed by the BT-structure of the interaction matrix. A matrix-vector multiplication can be transformed to a convolution, which is computed using a Fast Fourier Transform (FFT) technique in $\mathrm{O}(n\ln(n))$ operations (see Subsection 4.4). Note however, that alternative techniques exist that do not require a regular grid (see Subsection 4.5).

The BT-structure also permits acceleration of direct methods. Flatau *et al.* [122] used an algorithm for inversion of symmetric BT-matrices. It has complexity $\mathrm{O}(n^3/n_x)$ and storage requirements $\mathrm{O}(n^2/n_x)$, since only 2 block columns of the inverse matrix need to be stored. In this case the $x$-axis is oriented along the longest particle dimension. Recently Flatau [123] studied the special case of 1D DDA where all dipoles are located on a straight line and equally spaced, in which systems of equations for different components can be separated. The interaction matrix for each component is symmetric Toeplitz, and a modern fast algorithm can be applied for its inversion. This method requires preliminarily solving linear equations for two right sides (e.g., by some iterative technique); then multiplication of the inverse matrix by any vector (i.e., a solution of the linear system for any right part) requires only $\mathrm{O}(n\ln(n))$ operations. However, Flatau pointed out a strict limitation for all methods for fast calculation of the inverse of the interaction matrix: they are applicable only when polarizabilities of all dipoles are the same, since otherwise the first term on the right side of Eq. (20) ruins the BT structure on the diagonal of the interaction matrix. Therefore, they are currently limited to *homogenous rectangular* scatterers. Fortunately, it is not a problem for matrix-vector multiplication, since the diagonal term can be evaluated independently and added to the final result.

## 4.4 FFT

Goodman *et al.* [124] showed that multiplication of the interaction matrix for a rectangular lattice (see Subsection 4.3) by a vector can be transformed into a discrete convolution

$$\mathbf{y_i} = \sum_{j=1}^{N} \overline{\mathbf{G}}_{ij}\mathbf{P}_j = \sum_{\mathbf{j}=(1,1,1)}^{(n_x,n_y,n_z)} \overline{\mathbf{G}}'_{\mathbf{i-j}}\mathbf{P_j} = \sum_{\mathbf{j}=(1,1,1)}^{(2n_x,2n_y,2n_z)} \overline{\mathbf{G}}'_{\mathbf{i-j}}\mathbf{P}'_{\mathbf{j}}, \tag{87}$$

where $\overline{\mathbf{G}}'_{\mathbf{i}}$ is defined by Eq. (86) (and $\overline{\mathbf{G}}'_{\mathbf{0}} = 0$) for $\left|i_\mu\right| \le n_\mu$ and

$$\mathbf{P}'_{\mathbf{j}} = \begin{cases} \mathbf{P_j},\ \forall\mu:\ 1 \le j_\mu \le n_\mu \\ 0,\ \text{otherwise} \end{cases}. \tag{88}$$

Both $\overline{\mathbf{G}}'$ and $\mathbf{P}'$ are then regarded as periodic in each dimension μ with period $2n_\mu$. A discrete convolution can be transformed with a FFT to an element-wise product of two vectors, which is easily computed. It requires evaluation of a direct and inverse FFT for each matrix-vector product. Each of them is a 3D FFT of order $2n_x \times 2n_y \times 2n_z$. This operation is done for each of the 3 Cartesian components of $\mathbf{P}'$ and preliminary calculations is performed for 6 independent tensor components of $\overline{\mathbf{G}}'$.

A slightly different method can be devised based on the paper by Barrowes *et al.* [125], who developed an algorithm for multiplication of any MBT by a vector. The multiplication is brought down to a 1D convolution that is evaluated by two 1D FFTs of order $(2n_x - 1)(2n_y - 1)(2n_z - 1)$. Flatau [123] proposed an algorithm of matrix-vector multiplication for BT interaction matrix (e.g. 1D DDA), which requires twice as many FFTs as the standard algorithm, but of order $n$ instead of $2n$. Although Flatau stated that an extension of this algorithm to the general 3D case is straightforward, it is at least not trivial and probably its complexity will scale the same as standard methods.

## 4.5 Fast multipole method

The Fast Multipole Method (FMM) was developed by Greengard and Rokhlin [126] for efficient evaluation of the potential and force fields in $N$-body simulations where all pairwise interactions of $N$ particles are computed. The FMM is based on truncated potential expansions [127]. It is also called a hierarchical tree method because particles are grouped together in a hierarchical way, and the interaction between single particles and this hierarchy of particle groups is calculated [128]. However, some researchers distinguish between single- and multilevel FMM [129,130]; only the latter is truly hierarchical. The FMM naturally fits the DDA, since the matrix-vector multiplication is actually computing the total field on each single dipole due to all other dipoles, as was noted by Hoekstra and Sloot [128]. The computational complexity of the FMM (see below) is similar to FFT-based methods (see Subsection 4.4), but it does not require any regularity of the grid, thus making it applicable to any scatterer. The drawback is that the FMM is conceptually more complex, making it much harder to code. Nonetheless, the FMM was implemented in the DDA by Rahola [9,127].

Error analysis is critical for the FMM, since the acceleration is obtained by using approximations, in contrast to exact FFT-based methods. Approximation parameters are chosen to keep an error, calculated according to some estimate, in certain bounds. The more exact the error estimate is, the less computations are required; thus, the faster the whole algorithm. Therefore, algorithm complexity is directly connected to error analysis [131].

Koc and Chew [129] described the application of multilevel FMM to the DDA. They used semi-empirical formulae to determine the number of terms in multipole series, and obtained $O(N)$ complexity. However, rigorous, close to exact, error analysis is still lacking for the FMM applied to the DDA. It will allow obtaining a real algorithm complexity with guaranteed accuracy. Such an analysis has been conducted for 2D acoustic scattering [130], and for light scattering formulated in terms of surface integrals [131]. In both cases the FMM was proven to have an asymptotic complexity $O(N\ln^2(N))$. Application of the FMM to surface-integrals formulation of light scattering was reviewed by Dembart and Yip [132].

Another problem of implementing the FMM is that it is completely dependent upon the exact form of the interaction potential $\overline{\mathbf{G}}_{ij}$. All manuscripts mentioned above deal with interaction between point dipoles, i.e. Eq. (14). If a more complex expression for $\overline{\mathbf{G}}_{ij}$ is used (e.g. IT), most of the FMM should be developed anew. This makes integration of the FMM and the DDA a formidable problem.

The FMM is a promising method to calculate light scattering by particles that cannot be mapped effectively on a rectangular grid; however, there is still space for improving its theory to make it more robust and guarantee certain accuracy.

The FMM is not the only hierarchical tree method available. For instance, a very intuitively simple method was proposed by Barnes and Hut [133,134]. Multipole expansions over the center of mass in gravitational computations are used, contrary to geometrical center in the FMM. It automatically eliminates the second term in the multipole expansion, and allows fast evaluation of monopole terms. Though this method is much simpler and clearer than the FMM, it has very little control over the errors that can be studied almost exclusively empirically. It can be applied to the DDA without significant increase in the total computational errors.[7]

---

[7] Hoekstra AG, unpublished results

An alternative approach was proposed by Ding and Tsang [135]. They studied scattering from trees and used a sparse matrix iterative approach. The interaction matrix is divided into a strong part, which accounts for interaction between nearby dipoles, and a complement weak part: $\mathbf{A} = \mathbf{A}^s + \mathbf{A}^w$. The strong part is sparse and therefore allows quick solution of the linear system. The weak part is a small correction that is accounted for iteratively:

$$\mathbf{A}^s \mathbf{x}^{(0)} = \mathbf{y}\,,\ \mathbf{A}^s \mathbf{x}^{(l+1)} = \mathbf{y} - \mathbf{A}^w \mathbf{x}^{(l)}\,. \tag{89}$$

The authors demonstrate potential of this approach for some test cases.

### 4.6 Orientation averaging and repeated calculations

In many physical applications, one is interested in optical properties of an ensemble of randomly oriented particles. When the concentration of particles is small, multiple scattering is negligible and the optical properties are obtained by averaging single-particle scattering over different particle orientations. More general problems, where particles are not identical or multiple scattering is significant, are not considered here.

Orientation averaging of any scattering property can be described as the integral over the Euler's orientation angles (including a probability distribution function if necessary), which is brought down to a sum by appropriate quadrature. The problem therefore consists in calculation of some scattering property for a set of different orientations of the same particle. The easiest way is to calculate it by solving sequentially and independently each problem from the set. However, the large size of this set calls for some means of reducing the calculations. This is especially relevant when the particle is asymmetric; hence, its optical properties are sensitive to particle orientation. Let us further assume for clarity that we are interested in the scattering matrix at a certain scattering angle. All the discussion for other scattering properties is analogous or even simpler.

Singham *et al.* [136] noted that the set of problems described above is physically equivalent to a fixed orientation of the particle and different incident and scattering directions. The latter are determined by transformation of the laboratory reference frame to the reference frame associated with the particle. The amplitude scattering matrix, and hence the Mueller matrix, also is transformed along with the reference frame (see e.g. [137] for transformation formulae). There are two immediate advantages of using a fixed particle orientation. First, $\mathbf{A}$ is kept constant (see though discussion below), and therefore the construction of $\mathbf{A}$ is done only once. Second, the amplitude matrix for any scattering angle is quickly obtained after the linear system is solved (for two incident polarizations). Hence, integration over one Euler angle is relatively fast.

The constancy of **A** can be exploited to further reduce the time of orientation-averaging. If $\mathbf{A}^{-1}$ or its *LU* decomposition is obtained [75,136], a single solution for any right part **y** can be obtained in $n^2$ operations – the same or less time than required for *one* iteration using general iterative methods (see Subsection 4.1). Moreover, Singham *et al.* [136] and McClain and Ghoul [138] independently proposed an analytical way of averaging the scattering matrix at any scattering angle, which requires $O(n^2)$ operations once $\mathbf{A}^{-1}$ is known. Khlebtsov [139] extended this technique to averaging of extinction and absorption cross sections.

However, by employing special properties of the matrix **A** in the DDA allows computing matrix-vector products in $O(n\ln(n))$ operations (see Subsections 4.4, 4.5). Although some acceleration of direct methods also can be performed (see Subsection 4.3), they are still $O(n^2)$ or slower. For large $n$, iterative methods (assuming that they converge in much less than $n$ iterations) are clearly preferable, even if many quadrature points are used. Moreover, large $n$ is unattainable by direct methods because of storage requirements. Another improvement could be using a heavy preconditioner, which has large initialization cost and greatly increases convergence rate. Initialization cost is then justified because it is computed only once. Possible candidates are incomplete factorization preconditioners [107].

Above it was stated that **A** is constant for a fixed-orientation particle. However, modern DDA formulations (e.g. LDR) take into account the direction of light incidence. Hence **A** depends upon this direction, but only weakly through $O\left((kd)^2\right)$ corrections. This complicates the techniques described above, however probably they still may be used together with some special methods to correct for small changes in **A** on every step. Such methods have not been developed as yet.

Another possibility to perform orientation averaging is to first compute the T-matrix of the particle, which then allows analytical averaging [140]. The T-matrix formalism is based on the multipole expansion which is truncated at some order $N_0$. Although $N_0$ is hard to deduce *a priori*, usually it is several times $x$ [141,142]. The number of rows in the T-matrix equals $2N_0(N_0+2)$. The simplest way to evaluate the T-matrix based on the DDA is to solve for every incident spherical wave (i.e. for each row of the T-matrix) independently [141]. Then the above discussion about optimizing this repeated calculation is relevant. Using iterative techniques with $N_{iter}$ number of iterations, computation time is $N_0^2 N\left[O\left(N_{iter}\ln(N)\right)+O\left(N_0^2\right)\right]$, where the first term in the sum is the time for solving the linear system, and the second one is the actual computation of the values in the row of the T-matrix. A new method to obtain the T-matrix from the DDA interaction matrix was proposed by Mackowski [141]. This requires two summations with computational time $\mathrm{O}\left(N_0^2 N \ln(N)\right)$ and $\mathrm{O}\left(N_0^4 N\right)$. Mackowski showed that for $x=5$ his method is an order of magnitude faster than the straightforward one.

Recently Muinonen and Zubko [143] have proposed a way to optimize ensemble averaging of DDA results over different sizes and refractive indices. It is based on calculating a "good guess" for the initial vector in the iterative solver using results of the calculations with similar parameters. Similar ideas can be used to optimize simulation of a set of slightly different shapes or orientational averaging.

Use of repeated calculations to increase the accuracy of DDA simulations was proposed recently by Yurkin *et al.* [63]. Several independent simulations with different discretization parameter were performed and results were extrapolated to the infinite discretization giving better accuracy than those of a single DDA simulation.

# 5 Comparison of the DDA to other methods

Hovenier *et al.* [144] compared the DDA, the Extended Boundary Condition Method (EBCM), and the Separation of Variables Method (SVM) for calculations of scattering by spheroids, finite cylinders and bispheres. Parameters of the problems were as follows: $m=1.5+0.01\mathrm{i}$, equivolume size parameter $x=5$, $y=0.6$. The angular dependencies of scattering matrix elements were calculated. The EBCM and SVM seemed to achieve an exact solution, and the DDA showed little errors, except for backscattering angles, where they were up to 10-20%.

Wriedt and Comberg [145] compared the DDA, EBCM, and Finite Difference Time Domain (FDTD) method for a cube with $m=1.33$, 1.5 and $x=2.9$, 4.9, 9.7. For $x=2.9$ and 4.9. The DDA and EBCM achieved good accuracy in calculation of scattering intensity angle dependence; the DDA was 2-5 times faster, but consumed 8-16 times more memory ($y$ was in the range 0.3-0.5). The FDTD had similar computational requirements as the DDA but was less accurate. For $x=9.7$ the DDA was the only one to achieve little errors within the given computational resources.

Comberg and Wriedt [146] compared the DDA, GMM (see Subsection 3.3) and the Generalized Multipole Technique (GMT) for clusters of a few spheres. A single sphere had $x$ in the range 4-20 and $m=1.33$, 1.5. All the methods managed to achieve good accuracy, but the GMM was one order of magnitude (and for large $x$ even several orders) faster than the other two. The DDA and GMT also were used to compute scattering by a cluster of two oblate spheroids with $x = 5$ and $m = 1.33$. The DDA was less accurate and consumed 4 times more memory, but was 6 times faster than the GMT.

Wriedt *et al*. [147] compared the DDA, FDTD, GMT, and Discrete Sources Method (DSM) for the calculation of light scattering by a Red Blood Cell (RBC) with $x = 35$ and $m = 1.06$. Accuracy of all methods was similar. The DDA and GMT showed similar calculation times; they were 7 times faster than the FDTD and 12 times slower than the DSM. It should be noted that the latter employed the axisymmetric property of RBC.

Recently Yurkin *et al.* [58] systematically compared the DDA and the FDTD for spheres with *m* from 1.02 to 2 and *x* from 10 to 100, depending on *m*. It was shown that numerical performance of the DDA is much more sensitive to the refractive index than that of the FDTD. Therefore, the DDA is preferable for small *m,* the FDTD for larger *m*. Cleary, the crossover point is not well defined and will depend on the details of the problem at hand as well as on the particular implementations of both methods.

The main advantage of the DDA is that it is one of the most general methods, having a very broad range of applicability, limited only by available computational power. The reverse of this advantage is that it has almost no means to use the symmetry of the scatterer. Thus the DDA is not able to compete with the EBCM for homogenous axisymmetric scatterers. For homogenous non-axisymmetric scatterers the DDA is competitive with the EBCM for single-particle orientation, but the latter allows much faster orientation averaging. The EBCM has little applicability to inhomogeneous scatterers, where the DDA can be applied without any changes. Comparison between the FDTD and the DDA suggests that the DDA is more suitable for small *m*. It also should be noted that the FDTD is even more general, being easily applicable to non-harmonic incident electric fields. Moreover, simulation of one pulse incident wave with the FDTD gives the solution for a complete spectrum of incident harmonic plane waves, but with a limitation on accuracy.

# 6 Concluding remarks

The DDA has been reviewed using a general framework based on the integral equation for the electric field. Although mainstream DDA algorithms as used in several production computer programs, has not changed significantly since 1994, many different improvements have been proposed since that time. Some of them do improve the accuracy or numerical performance of the DDA; however, they still wait for a wide acceptance. It seems that a critical mass of new improvements is building up, hopefully resulting in a next breakthrough in the field of the DDA.

In our opinion, future major improvements in the DDA computer implementations will be connected with one of the following:

1) Decreasing shape errors by implementing WD or similar techniques.
2) Improving polarizability and interaction terms by techniques that are still to be developed similar to IT and PEL.
3) Studying different preconditioners for the DDA interaction matrix, either trying some of the known types or developing one considering the special structure of the matrix.

Item (1) should improve the overall accuracy of the DDA, especially for cases where shape errors are dominant, item (2) should expand the DDA applicability region to higher refractive indices, and item (3) should boost overall performance, especially for large size parameters and/or refractive indices.

## Acknowledgements

We thank Dan Mackowski for clarifying discussion on the simulations of scattering by clusters of spheres and Gorden Videen for critically reading the manuscript and for valuable discussions. Our research is supported by Siberian Branch of the Russian Academy of Sciences through the grant 2006-03.

## Appendix: description of used acronyms and symbols

<<No text here, only Table 2 and Table 3>>

## Tables

Table 1. Accuracy of different DDA formulations for a sphere.[a]

| Value | Method | $x$ | $a/d$ | $y$ | $m$ | Error, % | Ref. |
|---|---|---|---|---|---|---|---|
| $C_{ext}$ | $a_1$-term | 1÷2 | 2÷4[c] | 0.65<br>0.85 | 1.33+0.05i<br>1.7+0.1i | 3<br>6 | [38] |
| CSec, $S_{11}$ | LAK | 9<br>9<br>5 | 21[c]<br>29[c]<br>28[c] | 0.44<br>0.42<br>0.51 | 1.05<br>1.33+0.01i<br>2.5+1.4i | 0.05, 37<br>0.5, 35<br>4, 15 | [56] |
| $C_{sca}$, $C_{abs}$ | DGF | ≤3.2[c] | 16 | ≤1 | 4+3i | 5, 10÷30 | [3] |
| CSec | LDR | ≤8[c] | 16 | ≤0.5, ≤0.1 | m-1≤1 | 1, 2 | [3] |
| $C_{sca}$<br>$C_{abs}$ | LDR | ≤7[c] | 16 | ≤1<br>≤0.5, ≤1 | 2+i | 1.5<br>3, 4 | [3] |
| CSec | LDR | ≤16[c]<br>≤10[c] | 25 | ≤1 | 1.6+0.0008i<br>2.5+0.02i | 10<br>20 | [51] |
| CSec<br>$S_{11}$ | LDR | any | any | ≤1 | $\vert m\vert\le 3$ | 5<br>20÷30 | [4][e] |
| $C_{sca}$ | LDR | ≤10[c] | 16 | $kd$≤0.63 | 0.69<br>0.41<br>0.29 | 0.3<br>1<br>5 | [148] |
| $S_{11}$ | LDR | ≤10[c] | 24 | $kd$≤0.42 | 0.69<br>0.41 | 50<br>20 | [148] |
| $C_{ext}$, $S_{11}^{RMS}$ | LDR | 20÷160<br>20÷130<br>20÷60<br>20÷30<br>20÷30<br>20 | 32÷256[c]<br>40÷256[c]<br>48÷128[c]<br>56÷80[c]<br>64÷88[c]<br>64[c] | 0.61÷0.65[d]<br>0.56÷0.64[d]<br>0.58÷0.65[d]<br>0.57÷0.60[d]<br>0.56÷0.62[d]<br>0.62[d] | 1.05<br>1.2<br>1.4<br>1.6<br>1.8<br>2 | 0.04, 38<br>0.4, 23<br>1, 59<br>4.4, 56<br>5.7, 105<br>2.0, 86 | [113] |
| Ψ | FCD | π, 2π | 2.8, 5.6[c] | 1.7 | 1.5 | 1 | [44] |
| Ψ | WD-FCD | 0.5÷3.2[c]<br>1.5÷3.8[c]<br>0.9÷1.5[c] | 5<br>6<br>6 | $y^{Re}$=0.63 | $\vert m\vert<7$[b]<br>$\vert m\vert<2.5$[b]<br>$\vert m\vert<4$[b] | 0.1<br>0.1<br>1 | [10] |
| CSec<br>$C_{abs}$<br>$C_{ext}$ | IT | ≤5.2[c]<br>≤2.1[c]<br>≤1.1[c] | 8 | ≤1 | 1.5+0.3i<br>3.5+1.4i<br>7.1+0.7i | 2<br>20<br>15 | [11] |
| CSec | RCB-RR | ≤8.2[c]<br>≤7.5[c]<br>≤5.9[c]<br>≤3.4[c]<br>≤1.3[c] | 16 | ≤1 | 1.8+0.4i<br>1.9+i<br>2.5+i<br>2.5+4i<br>7.4+9.4i | 1<br>2<br>2<br>10<br>20 | [48] |
| CSec | SCLDR<br>SCLDR<br>RCB | ≤7.2[c]<br>≤1.5[c]<br>≤1.5[c] | 12 | ≤0.8 | 1.33+0.1i<br>5+4i<br>5+4i | 2<br>5<br>7 | [47] |

[a] All errors are relative. CSec denotes the maximum error over all cross sections, $S_{11}$ and $S_{11}^{RMS}$ correspond to maximum and root mean square error over the range of scattering angles, Ψ is the normalized mean error of the far-field electric fields [44]. In some cases two errors are shown in one cell separated by a coma. They correspond to two values of one of the parameters in the same row.

[b] approximate description of the range.

[c] this value is determined by other values in the same row.

[d] this value is slightly different for different size parameters.

[e] this corresponds to the “rule of thumb” for spheres.

Table 2. Acronyms in alphabetical order.

| Acronym | Description | Section[a] |
|---|---|---|
| (L) | Left Jacobi preconditioner | 4.1 |
| (R) | Right Jacobi preconditioner | 4.1 |
| $a_1$-term (M) | dipole term in the Mie theory | 3.1 |
| Bi-CGSTAB | Bi-CG Stabilized | 4.1 |
| BT | Block-Toeplitz | 4.3 |
| CEMD | Coupled Electric and Magnetic Dipole | 3.4 |
| CG | Conjugate Gradient | 4.1 |
| CGNR | CG applied to Normalized equation with minimization of Residual norm | 4.1 |
| CGS | CG Squared | 4.1 |
| CS | Complex Symmetric | 4.1 |
| CLDR | Corrected LDR | 3.1 |
| CM | Clausius-Mossotti | 2 |
| DDA | Discrete Dipole Approximation | 1 |
| DGF | Digitized Green's Function | 1 |
| DSM | Discrete Sources Method | 5 |
| EBCM | Extended Boundary Condition Method | 5 |
| FCD | Filtered Coupled Dipoles | 3.1 |
| FDTD | Finite Difference Time Domain | 5 |
| FFT | Fast Fourier Transform | 4.4 |
| FMM | Fast Multipole Method | 4.5 |
| GMM | Generalized Multiparticle Mie solution | 3.3 |
| GMRES | Generalized Minimal Residual | 4.1 |
| GMT | Generalized Multipole Technique | 5 |
| GPBi-CG | Generalized Product-type methods based on Bi-CG | 4.1 |
| IT | Integration of Green's Tensor | 3.1 |
| LAK | Lakhtakia | 3.1 |
| LC | Lou and Charalampopoulos | 3.3 |
| LDR | Lattice Dispersion Relation | 3.1 |
| MBT | Multilevel BT | 4.3 |
| MoM | Method of Moments | 3.1 |
| PEL | Peltoniemi | 3.1 |
| PP | Purcell and Pennypacker | 1 |
| QMR | Quasi-Minimal Residual | 4.1 |
| RBC | Red Blood Cell | 5 |
| RCB | Rahmani, Chaumet, and Bryant | 3.1 |
| RGD | Rayleigh-Gans-Debye approximation | 4.2 |
| RR | Radiative Reaction correction | 3.1 |
| SCLDR | Surface Corrected LDR | 3.1 |
| SOF | Scattering Order Formulation | 4.2 |
| SVM | Separation of Variables Method | 5 |
| TFQMR | Transpose Free QMR | 4.1 |
| WD | Weighted Discretization | 3.1 |

[a] where it is explained or first appears (if no explanation is given).

Table 3. Symbols used, Latin and Greek letters in alphabetical order.[a]

| Symbols | Description | Section |
|---|---|---|
| (0) | *superscript*: approximate value (usually under constant field assumption) | 2 |
| ($n$) | *superscript*: after $n$-th iteration | 4.2 |
| $\langle \cos\theta \rangle$ | asymmetry parameter | 2 |
| * | *superscript*: complex conjugate | 2 |
| $\mathbf{A}$ | a matrix | 3.1 |
| $\mathbf{A}$ | $\mathbf{k}/k$ | 2 |
| $a$ | radius of (equivalent) sphere | 3.1 |
| $\mathbf{B}$ | correction matrix in SCLDR | 3.1, Eq. (64) |
| $b_1 - b_3$ | numerical coefficients in polarization prescriptions | 3.1 |
| $\overline{\mathbf{C}}$ | tensor of electrostatic solution | 3.1, Eq. (60) |
| $C_{sca}$, $C_{abs}$, $C_{ext}$ | Scattering, absorption, extinction cross section | 2 |
| $c$ | speed of light in vacuum | 2 |
| $d$ | size of a cubical cell | 2 |
| $\mathbf{E}$, $\mathbf{E}^{inc}$, $\mathbf{E}^{exc}$, $\mathbf{E}^{self}$, $\mathbf{E}^{sca}$ | (total) electric field, incident, exciting, self-induced, scattered | 2 |
| $\mathbf{e}^0$ | polarization vector of the incident wave, $\lvert\mathbf{e}^0\rvert = 1$ | 2 |
| $e$ | *superscript*: effective | 3.1 |
| $\mathbf{F}$ | scattering amplitude | 2 |
| $f$ | a function;<br>volume filling factor | 3.1<br>3.1 |
| $\overline{\mathbf{G}}$ | free space dyadic Green's function (tensor) | 2 |
| $\overline{\mathbf{G}}^s$ | $\overline{\mathbf{G}}$ in static limit | 2 |
| $\overline{\mathbf{G}}_{ij}$ | interaction term | 2 |
| $H$ | *superscript*: conjugate transpose | 3.1 |
| $h^r$ | impulse response function of a filter | 3.1 |
| $\overline{\mathbf{I}}$, $\mathbf{I}$ | identity dyadic (tensor), operator (matrix) | 2, 4.1 |
| $\mathbf{i}$, $\mathbf{j}$ | *subscript*: vector indices | 4.3 |
| i | imaginary unity | 2 |
| $i$, $j$ | *subscript*: number of the dipole | 2 |
| $K$ | order of a BT matrix | 4.3 |
| $\mathbf{k}$ | free space wave vector | 2 |
| $\overline{\mathbf{L}}$ | self-term dyadic | 2 |
| $\mathbf{M}$ | integral associated with finiteness of $V_0$;<br>preconditioner | 2<br>4.1 |
| $\overline{\mathbf{M}}$ | dyadic associated with $\mathbf{M}$ | 2 |
| $m$ | refractive index (relative) | 3.1 |
| $N$ | total number of dipoles | 2 |
| $\mathbf{n}$ | $\mathbf{r}/r$ | 2 |
| $n$ | size of a matrix | 4.1 |
| $\hat{n}'$ | external normal to the surface | 2 |
| $n_x$, $n_y$, $n_z$ | sizes of the rectangular lattice | 4.3 |
| $\mathbf{P}$ | polarization | 2 |
| $p$ | *superscript*: principal | 3.1 |
| $q$ | $2\pi/d$ | 3.1 |

| $\mathbf{R}$ | $\mathbf{r}-\mathbf{r}'$ | 2 |
|---|---|---|
| $R_0$ | radius of the smallest sphere circumscribing the scatterer | 4.2 |
| $\mathbf{r}$, $\mathbf{r}'$ | radius-vectors | 2 |
| $S$ | LDR coefficient dependent on incident polarization | 3.1, Eq. (49) |
| $S_i$ | amplitude matrix element | 3.2 |
| $S_{ij}$ | Mueller matrix element | 3.2 |
| $s$ | *superscript*: secondary;<br>strong;<br>*subscript*: equivalent spherical dipole | 3.1<br>4.5<br>3.1 |
| $\overline{\mathbf{T}}$ | boundary condition tensor | 3.1, Eq. (66) |
| $t$ | time | 2 |
| $V$ | volume of the scatterer | 2 |
| $V_0$ | exclusion volume | 2 |
| $w$ | *superscript*: weak | 4.5 |
| $\mathbf{x}$ | unknown vector | 4.1 |
| $x$ | size parameter of scatterer | 3.2 |
| $x$, $y$, $z$ | Cartesian coordinates | 4.3 |
| $\mathbf{y}$ | a known vector (right side of a linear system) | 4.1 |
| $y$ | $\lvert m \rvert kd$ | 3.2 |
| $y^{\mathrm{Re}}$ | $\mathrm{Re}(m)kd$ | 3.2 |
| $\alpha$, $\overline{\boldsymbol{\alpha}}$ | polarizability, tensor | 2 |
| $\gamma$ | optimal reduction factor | 4.1 |
| $\delta$ | Kronecker symbol | 3.1 |
| $\varepsilon$ | electric permittivity (relative) | 2 |
| $\eta$ | correction function | 3.3 |
| $\overline{\mathbf{\Lambda}}$ | intermediate tensor in RCB method | 3.1, Eq. (62) |
| $\mathbf{\Lambda}$ | linear integral operator, its matrix | 4.2 |
| $\mu$, $\nu$, $\rho$, $\tau$, … | *sub-, superscript*: Cartesian components of vectors (tensors) | 2 |
| $\xi$, $\psi$ | Riccati-Bessel functions | 3.1 |
| $\chi$ | electric susceptibility | 2 |
| $\Psi$ | mean relative error of far-field electric field | 3.2 |
| $\Omega$ | solid angle | 2 |
| $\omega$ | circular frequency of the harmonic electric field | 2 |

[a] common sub- and superscripts are given on their own. For all vectors – the same symbol but in italic (instead of bold) denotes Euclidian norm of the vector (except unitary vectors).

## Figures

Fig. 1. Scheme of interrelation between the different DDA models discussed in Section 3.1. Arrows down correspond to assumptions employed. Vertical position of the method qualitatively corresponds to its accuracy (higher = better), however methods in different columns cannot be compared directly.

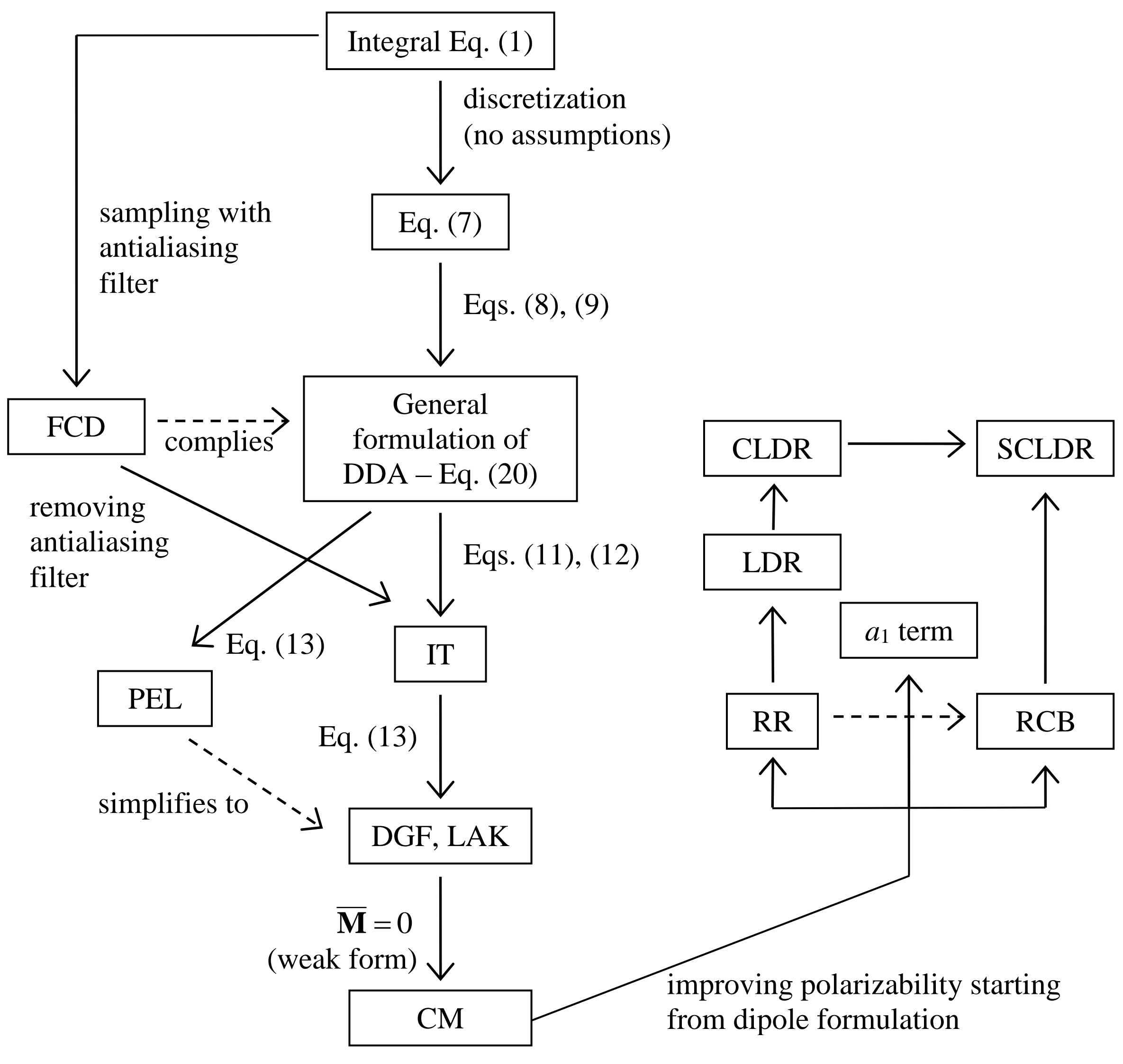

Fig. 1 Yurkin